\documentclass[12pt, a4paper]{article}

\usepackage[
  margin=0.75in,
  headsep=10pt, 
]{geometry}

\usepackage{graphicx}
\usepackage{soul}
\usepackage{graphicx}
\usepackage{epstopdf, epsfig}
\usepackage{amsmath}
\usepackage[table]{xcolor}
\usepackage{subcaption}
\usepackage{soul}
\usepackage{makecell}
\usepackage{multirow}
\usepackage{breqn}
\usepackage{amssymb} 
\usepackage{booktabs}
\usepackage{dcolumn}
\usepackage{bm}
\usepackage[utf8]{inputenc}
\usepackage[T1]{fontenc}
\usepackage{mathptmx}
\usepackage{etoolbox}
\usepackage{hyperref}
\usepackage{subcaption}
\usepackage{ragged2e}
\usepackage[sort&compress]{natbib} 
\usepackage[utf8]{inputenc}
\usepackage[T1]{fontenc}
\usepackage{graphicx}
\usepackage{epstopdf, epsfig}
\usepackage{xcolor}
\usepackage[table]{xcolor}
\usepackage{soul}
\usepackage{makecell}
\usepackage{multirow}
\usepackage{booktabs}
\usepackage{dcolumn}

\newcommand{\edt}[1]{{\color{black}#1}} 
\newcommand{\bi}[1]{{\color{black}#1}} 
\newcommand{\fm}[1]{{\color{black}#1}} 

\newcommand{\fmrev}[1]{{\color{black}#1}}

\usepackage{amsmath}   
\usepackage{amssymb}   
\usepackage[normalem]{ulem}

\newcommand{\soptitle}{From Dead-Band to Steady Rotation: Three-Dimensional Vortex Dynamics of Self-Starting Vertical-Axis Wind Turbines}
\usepackage{xcolor}

\begin{document}

\begin{center}
\Large \bf{\soptitle}
\vspace{0.1in}
\end{center}

\begin{center}
{Faisal Muhammad$^{1}$, Basel Ismail$^{2}$, and Muhammad Saif Ullah Khalid$^{1,\ast}$}
\vspace{0.1in}
\end{center}
\begin{center}
$^1$Nature-Inspired Engineering Research Lab (NIERL), Department of Mechanical and Mechatronics Engineering, Lakehead University, Thunder Bay, ON P7B 5E1, Canada\\
$^2$Department of Mechanical and Mechatronics Engineering, Lakehead University, Thunder Bay, ON P7B 5E1, Canada\\
\vspace{0.05in}
$^\star$\small{Corresponding Author, Email: mkhalid7@lakeheadu.ca}
\end{center} 

\begin{abstract}
The self-starting behavior of vertical-axis wind turbines is strongly influenced by the development of vortical structures along the span. However, a fully coupled high-fidelity three-dimensional analysis of this process requires substantial computational resources. This study presents a computationally efficient, spanwise-periodic, three-dimensional large-eddy simulation framework in OpenFOAM. The motion of the rotor is prescribed using a combination of two logistic functions to reproduce the angular-velocity evolution of a vertical-axis wind turbine undergoing self-starting. The results show that boundary-layer separation and the subsequent roll-up of the separated shear layer generate detached vortical structures that convect into the wake as finite spanwise segments. The deformation of these structures along the span increases with increasing spanwise extent. The dead-band is characterized by sustained wake interference and weak net acceleration, whereas the rapid-acceleration stage exhibits greater attachment of the flow, with separation confined to smaller regions near the blade. During quasi-steady operation, the shedding of vortices into the wake becomes repeatable, and downstream blade--vortex interactions are reduced. The approach based on the prescribed motion of the rotor enables computationally efficient, spanwise-resolved studies of vortex evolution suitable for parametric investigations and the evaluation of flow-control devices.
\end{abstract}

\section{Introduction}
\label{sec:Intro}
The increasing demand for renewable energy \edt{motivates} researchers to broaden the search for sustainable energy resources. In this context, wind energy, particularly vertical-axis wind turbines (VAWTs), \edt{attracts} growing attention \bi{due to key structural and aerodynamic benefits}. \bi{Beyond} easier maintenance \bi{from housing} heavy components near the ground, \bi{VAWTs exhibit directionally invariant aerodynamic operation, eliminating the need for complex yaw-tracking systems. They also perform effectively in turbulent airflow, lower structural loads by reducing top-heavy mass, and enable closer turbine spacing within wind farms to maximize power generation per unit of land.} However, these benefits are often overshadowed by poor \edt{aerodynamic} performance \bi{at} low \edt{wind speeds}, where complex unsteady \bi{phenomena, most notably dynamic stall,} can hinder the self-starting process. \bi{Developing a clear understanding requires careful examination of every phase, from initial motion to} quasi-steady rotation. \bi{Because these transient phases occur rapidly, capturing real-time unsteady flow variations using current experimental techniques remains extremely challenging. It is particularly true near the fast-moving blades, where rapid changes in the boundary layer, transient vortex shedding, and optical reflections of the blades' surfaces severely restrict high-resolution spatial and temporal measurements.} Hence, A \edt{better} understanding of self-starting \edt{of VAWTs can be obtained through high-fidelity} three-dimensional ($\mbox{3D}$) computational simulations. Nevertheless, much of the available literature on self-starting relies on two-dimensional ($\mbox{2D}$) \edt{flow} analysis. While $\mbox{3D}$ simulations are computationally expensive, they can provide \edt{more insights for} formation, growth, and decay of vortical structures generated by the motion of the blades and \edt{through underlying blade-vortex interactions. Nonetheless,} $\mbox{2D}$ \edt{simulations are} valuable for isolating and interpreting the underlying physics of self-starting and for guiding the design of more targeted $\mbox{3D}$ investigations. As noted by Bianchini et al. \cite{bianchini2017effectiveness}, when key parameters associated with the mesh \bi{(such as grid refinement and near-wall resolution)} and numerical modeling \bi{(such as selection of a turbulence model, size of the time-step, and spatial discretization schemes)} are appropriately selected, $\mbox{2D}$ computational fluid dynamics ($\mbox{CFD}$) simulations can show good agreement with \edt{experimentally determined} trends \edt{for the performance-related metrics of VAWTs}.

\fmrev{Most of the \edt{existing} literature related to self-starting of VAWTs is limited to \edt{research based on} $\mbox{2D}$ \edt{computational simulations}. These investigations primarily focused on geometric modifications to the rotor to improve its self-starting capability. Asr et al. \cite{asr2016study} investigated H-Darrieus turbines equipped with different NACA four-digit airfoils and found that a medium-thickness cambered NACA2418 airfoil with an outward pitch angle of $1.5^{\circ}$ provided the most favourable start-up characteristics, reducing the start-up time without compromising the peak performance \edt{of the turbine}. Arab et al. \cite{arab2017numerical} simulated \edt{flows around a} turbine\edt{, starting} from rest until it reached a steady rotational state while \edt{varying} the rotor's moment of inertia. Their results showed that the instantaneous aerodynamic torque must be considered when self-starting behavior is \edt{examined}. Celik et al. \cite{celik2020aerodynamic} subsequently identified a drag-assisted regime when the turbine \edt{rotated} with tip-speed ratio less than unity, followed by a fully lift-driven regime after the tip-speed ratio exceeded unity. They further showed that increasing the moment of inertia prolonged the time required to reach the final rotational speed and reduced its oscillations, without noticeably affecting the ability to start or the final speed. Increasing the number of blades shortened the start-up time but reduced the peak power coefficient, thus revealing a trade-off between self-starting and performance related to the power of the rotor. Ramírez and Saravia \cite{ramirez2021assessment} evaluated the suitability of the Reynolds-averaged Navier-Stokes \edt{equations (URANS) based} approach for modeling the start-up regime and identified portions of the rotor\edt{'s} revolution in which the predicted aerodynamic coefficients became unreliable, emphasizing the sensitivity of the transient response to the modeling of separated and transitional flows. Mohamed et al. \cite{mohamed2021better} investigated the flow physics governing start-up and confirmed that drag \edt{contributed} substantially to \edt{production of torque} during the initial cycles \bi{until} the tip-speed ratio reaches approximately $1.5$, whereas lift becomes increasingly dominant as the rotor accelerates. Their results also demonstrated that blade-\edt{vortex} interactions \edt{altered} the aerodynamic loading over particular downstream azimuthal positions. \edt{Additionally,} Khalid et al. \cite{khalid2022self} compared the flow-induced responses of single- and dual-stage Darrieus turbines and showed that the relative orientation of the two stages strongly \edt{affected} their torque generation and self-starting behavior, with appropriate staging improving the rotational response. Xu et al. \cite{xu2024study} combined wind tunnel\edt{-based} measurements with $\mbox{2D}$ simulations \edt{of passive flow-induced rotations of turbines} and demonstrated that a positive static torque \edt{might} initiate rotation but \edt{did} not necessarily ensure completion of the self-starting process. They found that the NACA2418 rotor exhibited better self-starting and \edt{power production} than the NACA0018 rotor\edt{. They also noted} that increasing the pitch angle from $0^{\circ}$ to $5^{\circ}$ and $10^{\circ}$ reduced its start-up time by 12\% and 20\%, respectively. Zare Chavoshi and Ebrahimi \cite{zare2024self} applied plasma actuation as an active flow-control strategy and reported that it converted the negative torque generated at low tip-speed ratios into positive torque, increased the tip-speed-ratio growth by up to 8\%, and shortened the overall self-starting period. \bi{Abul-Ela et al. \cite{abul2025enhancing} employed transient $\mbox{2D}$ CFD simulations to examine the effects of airfoil profile, camber orientation, and pitch angle on the self-starting behavior of VAWTs. Their results showed that appropriate modification of these geometric parameters improved the starting torque and reduced the self-starting time. Chen et al. \cite{chen2025improvement} investigated the use of blade spoilers and modifications to the inner and outer rotors of a hybrid VAWT using $\mbox{2D}$ CFD simulations and reported an improvement in its self-starting torque.} More recently, Erkan et al. \cite{erkan2025taguchi} extended the investigation of self-starting from an isolated rotor to \edt{farms of VAWTs} using $\mbox{2D}$ simulations. Although these studies provided important insight into the transient acceleration of Darrieus rotors, their predominantly $\mbox{2D}$ formulations cannot reproduce spanwise \edt{developments in the flow}, \edt{losses from tips of blades}, or the evolution of $\mbox{3D}$ vortex structures, \edt{governing their aerodynamic performance}.}

\fmrev{Although \edt{their exists a vast body of knowledge about aerodynamics of VAWTs through $\mbox{3D}$ simulations, a wide majority of them are based on prescribed rotations with fixed angular velocities. Some exampled include the work on dynamic stall experienced by the blades \cite{orlandi20153d,gosselin2016parametric}, spanwise variations in the flow \cite{li20132,escudero2024vorticity}, and influence of finitely-spanned blades \cite{alaimo20153d,franchina2019three,franchina2020three}. To the authors' knowledge, there exists two studies \cite{liu2026comprehensive, fatahian2024optimization} that reported $\mbox{3D}$ simulations for passive rotations of VAWTs. Very recently,} Liu et al. \cite{liu2026comprehensive} employed LES to assess the effects of \edt{camber of blades} and \edt{their} pitch angle on self-starting \edt{of a VAWT} using the initial moment coefficient, static moment coefficient, and start-up speed. They compared the flow structures around selected configurations and examined the temporal evolution of the wake using modal decomposition. \edt{However, their analysis on vortex dynamics was more focused on explaining differences between geometric configurations, which did not provide any information about how vortices are evolved and interact with the blades during different phases of the whole process. Besides, the work of Fatahian et al. \cite{fatahian2024optimization} proposed a dynamic 3D CFD model to assess the influence of the rotor's moment of inertia on the self-starting of pairs of VAWTs. However, this research was aimed at performing optimizing the spatial arrangement of VAWTs without discussing important flow-related phenomena around the blades and in the wakes of turbines.}}

\fmrev{One of the main reasons \edt{behind scarce literature from} $\mbox{3D}$ investigations of the self-starting process \edt{of VAWTs} is the \edt{prohibitively} high computational cost associated \edt{with modeling flow-induced rotations of turbines}. To address this limitation, \edt{our} present study \edt{proposes} a kinematic approach based on the logistic function to reproduce the principal stages of the self-starting process. \edt{Previously,} such functions \edt{were} used in some studies to represent the nonlinear power curves of wind turbines \cite{villanueva2016reformulation,villanueva2018comparison,seo2019wind,tao2019integrated,virgolino2020gaussian,jing2021wind}. \edt{In this work, we further build} on \edt{characteristic of} logistic \edt{functions to model saturation-like processes, such as self-starting behavior of VAWTs. With this context and background, the main novelty of our current work is constituted by two important aspects: (i) modeling the kinematics of VAWTs using a combination of logistic functions to handle multiple phases involved in their self-starting process, and (ii) examination and explanation of vortex dynamics around the blades and their complex interactions through $\mbox{3D}$ large-eddy simulations.}}

\section{\edt{Computational Methodology}}
\label{sec:NumVal}

\subsection{Geometry and Domain}
\label{subsec:Geom}

\begin{figure}
  \centering
  \begin{subfigure}[t]{0.5\linewidth}
    \centering
    \includegraphics[width=\linewidth]{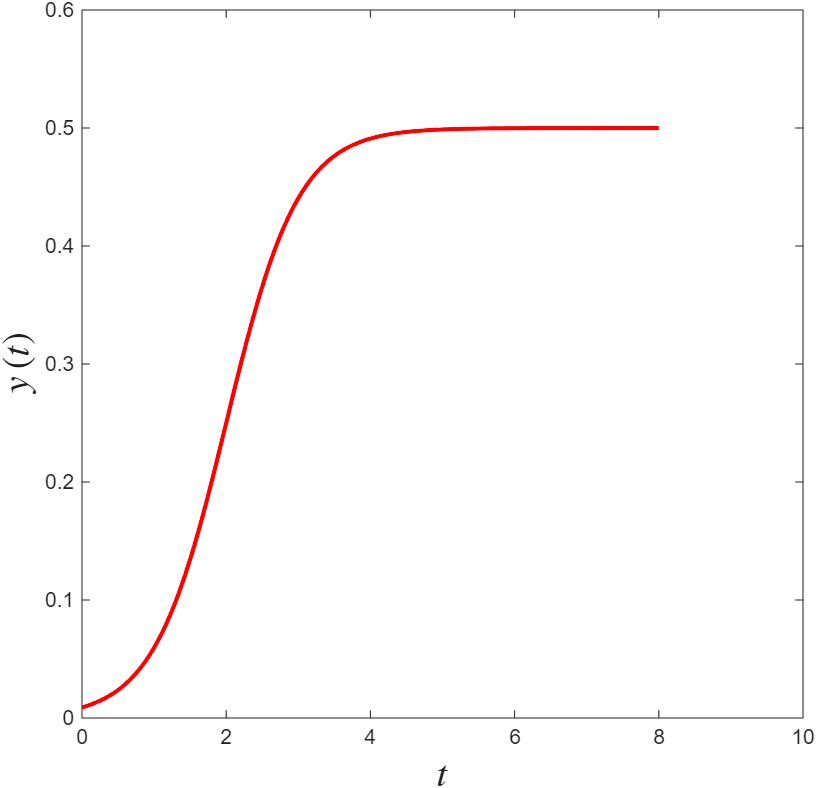}
    \caption{Baseline logistic function ($L=0.5$, $k=2$, $t_0=2$)}
    \label{fig:logisticmap}
  \end{subfigure}\hfill
  \begin{subfigure}[t]{0.5\linewidth}
    \centering
    \includegraphics[width=\linewidth]{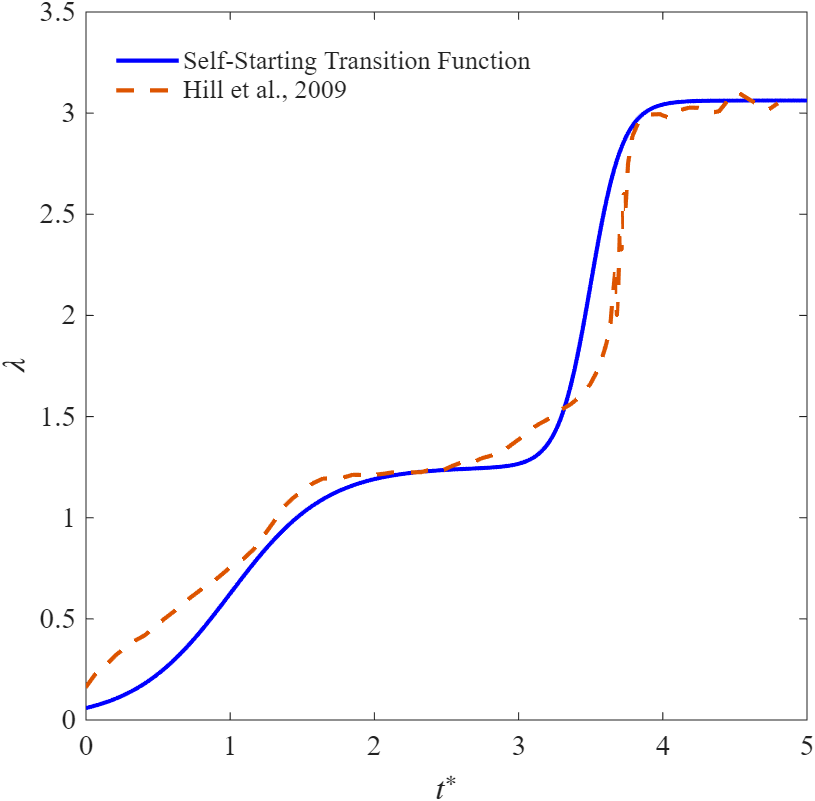}
    \caption{Proposed self-starting transition function compared with Hill et al.~\cite{hill2009darrieus}}
    \label{fig:selfstartfunc}
  \end{subfigure}
  \caption{Baseline logistic function and the prescribed self-starting transition used to define the motion of the rotor}
  \label{fig:logistic_vs_selfstart}
\end{figure}

The \edt{primary} aim of this work is to analyze \edt{vortex dynamics and blade-vortex interactions} during the self-starting process of VAWTs. Resolving the $\mbox{3D}$ flow field while calculating the flow-induced rotation of the rotor is computationally \edt{very} expensive, particularly when \edt{more advanced techniques, such as $\mbox{LES}$} is employed. To reduce the computational cost, \edt{we propose an effective way to determine} the time-dependent angular velocity of the rotor, \edt{(}$\omega(t)$\edt{)} to reproduce the principal stages of the self-starting process. The prescribed variation\edt{s} in $\omega(t)$ \edt{are} constructed using the logistic growth function, which is widely used to model transitions toward a saturation level \edt{in different fields of research} \cite{verhulst1844recherches}. The baseline logistic function, \edt{a plot of which is} shown in Fig.~\ref{fig:logisticmap}, is presented by Eq.~\ref{eq:log}\edt{:}

\begin{equation}
y(t)=\frac{L}{1+e^{-k\left(t-t_0\right)}}
\label{eq:log}
\end{equation}

\noindent where $L$ is the saturation level (carrying capacity), $k$ is the growth rate, and $t_0$ is the midpoint time. While Eq.~\ref{eq:log} captures a smooth transition between two states, a VAWT undergoing self-starting typically exhibits multiple stages, including the dead-band followed by a second stage of acceleration. Therefore, \edt{our present work involves superposition of} two logistic functions to define \edt{variations} in $\omega(t)$ of the rotor \edt{in a prescribed manner,} as shown in Fig.~\ref{fig:selfstartfunc} and expressed in Eq.~\ref{eq:wlogistic} \edt{ below:}

\begin{equation}
\omega(t)=\frac{L_1}{1+e^{-k_1\left(t-t_1\right)}}+\frac{L_2}{1+e^{-k_2\left(t-t_2\right)}}
\label{eq:wlogistic}
\end{equation}

\noindent where the first term represents the initial acceleration toward the dead-band, and the second term represents the subsequent rapid acceleration toward quasi-steady operation. The constants $L_1$, $k_1$, $t_1$ and $L_2$, $k_2$, $t_2$ are selected iteratively so that Eq.~\ref{eq:wlogistic} matches the qualitative self-starting behavior reported by Hill et al.~\cite{hill2009darrieus}. \edt{Particularly}, \edt{these} parameters are tuned to reproduce the principal stages of self-starting, namely the initial acceleration, the dead-band, rapid acceleration, and quasi-steady operation. The adopted values are $L_1=20$, $k_1=3$, $t_1=1$, $L_2=29$, $k_2=9$, and $t_2=3.5$. Next, the azimuthal position of the rotor \edt{(}$\theta(t)$\edt{)} is obtained by integrating the prescribed $\omega$ in Eq.~\ref{eq:wlogistic} \bi{over} time. Using

\begin{equation}
\int \frac{L}{1+e^{-k(t-t_0)}}\,dt
= Lt+\frac{L}{k}\ln\!\left(1+e^{-k(t-t_0)}\right)+C
\end{equation}

the resulting expression for $\theta(t)$ becomes
\begin{equation}
\theta(t)=49t+\frac{20}{3}\ln\!\left(1+e^{-3(t-1)}\right)
+\frac{29}{9}\ln\!\left(1+e^{-9(t-3.5)}\right)-121.8
\label{eq:thetalogistic}
\end{equation}

\noindent where the constant term \edt{is} chosen to satisfy the desired initial condition for the position of the rotor. \edt{Then, Eq.}~\ref{eq:thetalogistic} is implemented through a custom motion library to prescribe the rotation of the VAWT in \edt{the computational solver}.

\begin{figure}[h!]
    \centering
    \includegraphics[width=1\linewidth]{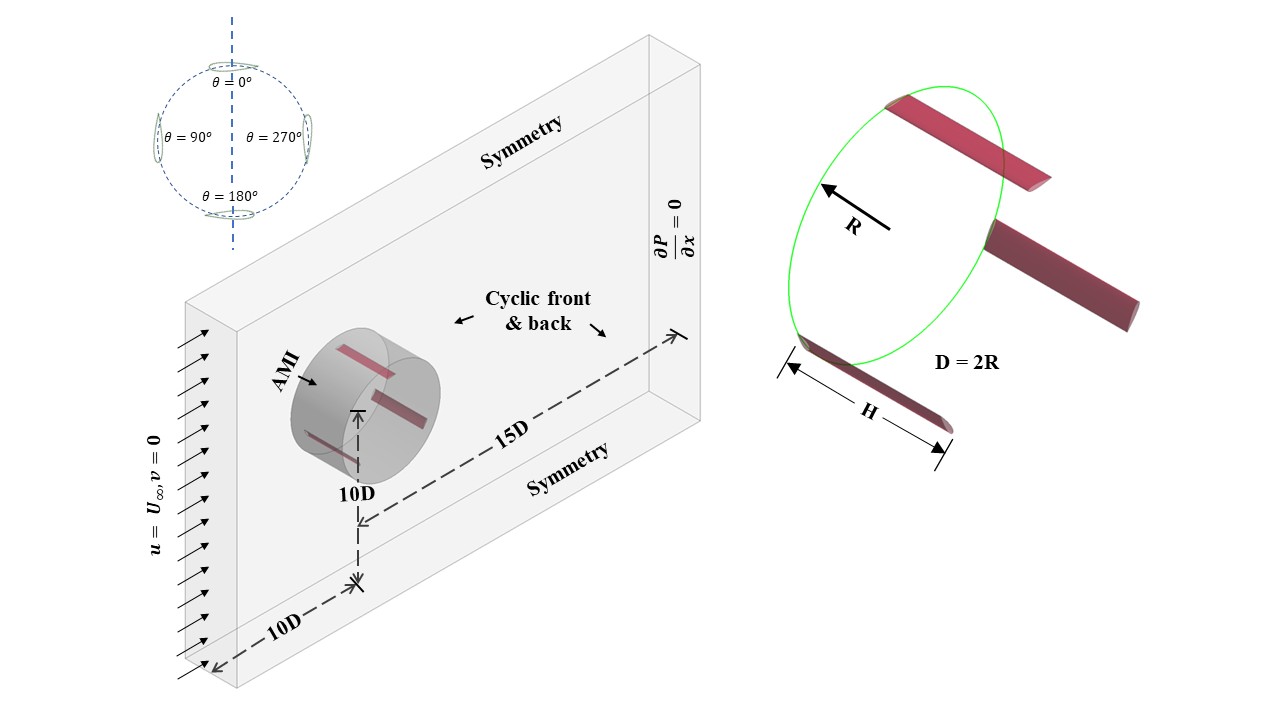}
    \caption{Schematic of the \edt{geometry of the} VAWT and computational domain (not to scale)}
    \label{fig:geometry_domain}
\end{figure}

Moreover, the transient response of self-starting reported by Hill et al. \cite{hill2009darrieus} (Fig.~\ref{fig:selfstartfunc}) is normalized in time so that the total duration of the proposed transition remains comparable with that of the reference response. This normalization preserves sufficient physical time for the formation, growth, and decay of the dominant vortical structures during each stage of start-up, thereby enabling meaningful identification and analysis of vortices within the present CFD framework. The geometry of the turbine\edt{, employed in our present work, also} follows the benchmark H-rotor configuration reported by Rainbird \cite{rainbird2007aerodynamic} and Hill et al. \cite{hill2009darrieus}, which \edt{was} subsequently adopted in numerous investigations of self-starting \edt{phenomena} as a reference case for validating numerical models \cite{sun2020effects,erkan2025taguchi,fatahian2024optimization}. The rotor consists of three straight blades that are uniformly spaced in azimuth and have symmetric NACA0018 sections. The chord length of each blade ($c$), the radius of the rotor ($R$), and the moment of inertia of the rotor ($J$) are $0.083~\mathrm{m}$, $0.375~\mathrm{m}$, and $0.018~\mathrm{kg\,m^2}$, respectively, as reported for the reference configuration. The height of the turbine, which corresponds to the spanwise length in the present model, is denoted by $H$ (Fig.~\ref{fig:geometry_domain}) and is varied to examine the evolution of vortices in the spanwise direction.

The computational domain is constructed to ensure that the inlet, outlet, and lateral boundaries do not interfere with the near-wake development of the rotor. As shown in Fig.~\ref{fig:geometry_domain}, the inlet plane is located \edt{at a distance of} $10D$ upstream of the center of the rotor and the outlet plane is located $15D$ downstream\edt{. T}he top and bottom boundaries are each placed $10D$ \edt{away} from the center of the rotor. These values \edt{are} selected based on the recommendations by Rezaeiha et al. \cite{rezaeiha2017cfd} to minimize blockage and boundary effects. A uniform free-stream velocity ($U_\infty$) is prescribed at the inlet, and a zero streamwise pressure-gradient condition is applied at the outlet. The top and bottom boundaries are treated as symmetry planes. The front and back boundaries are treated as cyclic planes so that each modeled height represents a spanwise-periodic configuration. This treatment excludes physical end effects while permitting three-dimensional vortex evolution along the axis of the turbine.

Rotation of the turbine is handled using a sliding-mesh formulation. A cylindrical rotating sub-domain enclosing the blades is embedded within a stationary far-field region, and the two regions communicate through an arbitrary mesh interface (AMI) to maintain conservative flux transfer across the sliding interface (Fig.~\ref{fig:geometry_domain}).

\subsection{Governing Equations for Flow Dynamics}
\label{subsec:CM}

The prescribed self-starting kinematics are implemented in OpenFOAM using a custom motion library that imposes $\theta(t)$ and computes the corresponding $\omega(t)$, while LES is used to resolve the flow field. In incompressible LES, the spatially filtered governing equations in Cartesian coordinates $(x,y,z)$ are \cite{sagaut2006large,pope2001turbulent}\edt{.}

\begin{align}
\frac{\partial \bar{u}_i}{\partial x_i} &= 0,
\qquad i\in\{x,y,z\},
\label{eq:lesCont}
\\
\frac{\partial \bar{u}_i}{\partial t}
+ \frac{\partial (\bar{u}_i\bar{u}_j)}{\partial x_j}
&=
-\frac{1}{\rho}\frac{\partial \bar{p}}{\partial x_i}
+ \nu\frac{\partial^2 \bar{u}_i}{\partial x_j\partial x_j}
-\frac{\partial \tau_{ij}}{\partial x_j},
\qquad i,j\in\{x,y,z\},
\label{eq:lesMom}
\end{align}

\noindent where $\bar{(\cdot)}$ denotes a filtered quantity, $\rho$ is the fluid density, $\nu$ is the kinematic viscosity, and $\tau_{ij}$ is the subgrid-scale (SGS) stress tensor\edt{.}

\begin{equation}
\tau_{ij}=\overline{u_i u_j}-\bar{u}_i\bar{u}_j.
\label{eq:sgsStressDef}
\end{equation}

\edt{Here,} the deviatoric SGS stress is closed using an eddy-viscosity assumption\edt{:}

\begin{equation}
\tau_{ij}-\frac{1}{3}\tau_{kk}\delta_{ij} = -2\nu_t \bar{S}_{ij},
\qquad
\bar{S}_{ij}=\frac{1}{2}\left(\frac{\partial \bar{u}_i}{\partial x_j}+\frac{\partial \bar{u}_j}{\partial x_i}\right),
\label{eq:boussinesq}
\end{equation}

\noindent where $\nu_t$ is the SGS eddy viscosity. \edt{It} is obtained using the Wall-Adapting Local Eddy-Viscosity (WALE) model. In compact form, the WALE model is written as \cite{ducros1998wall}\edt{:}

\begin{equation}
\nu_t = (C_w\Delta)^2\,
\frac{\left(S^{d}_{ij}S^{d}_{ij}\right)^{3/2}}
{\left(\bar{S}_{ij}\bar{S}_{ij}\right)^{5/2}+\left(S^{d}_{ij}S^{d}_{ij}\right)^{5/4}},
\label{eq:wale}
\end{equation}

\noindent where $\Delta$ is \edt{width of} the LES filter\edt{,} and $S^{d}_{ij}$ is defined from the resolved velocity-gradient tensor $g_{ij}=\partial \bar{u}_i/\partial x_j$ as\edt{:}

\begin{equation}
S^{d}_{ij}=\frac{1}{2}\left(g_{ik}g_{kj}+g_{jk}g_{ki}\right)-\frac{1}{3}\delta_{ij}\,g_{mn}g_{nm}.
\label{eq:waleSd}
\end{equation}

Near-wall effects on \edt{blades of} the turbine are treated using a wall-function, where the wall shear stress\edt{es are} obtained using a logarithmic law-of-the-wall relation.

\edt{In our present simulations, a} second-order implicit backward scheme is used for temporal discretization to capture self-starting with improved stability. The convective term in the momentum equation is discretized using a linear-upwind formulation with gradient reconstruction, providing an upwind bias for robustness while maintaining good accuracy in smooth regions. Convective transport of turbulence variables is handled using a limited linear scheme to preserve boundedness during the strongly unsteady start-up phases.

\subsection{Numerical Sensitivity and Validation}
\label{subsec:Space}

The grid and time-step independence studies are performed sequentially following the procedure reported by Li et al. \cite{li20132}. Grid convergence is first assessed using the 2D flow-induced rotation framework with URANS and $k$--$\omega$ shear-stress transport (SST) turbulence modeling. This assessment is used to identify an in-plane mesh that \edt{makes the numerical solution} sufficiently insensitive to further refinements \edt{in the grid}. \edt{Consequently,} the selected medium mesh is used for the time-step independence study. After the spatial and temporal resolutions are selected, the 2D mesh is extruded in the spanwise direction to construct the 3D computational grids used in the present simulations.

\begin{figure}
  \centering
  \includegraphics[width=1\linewidth]{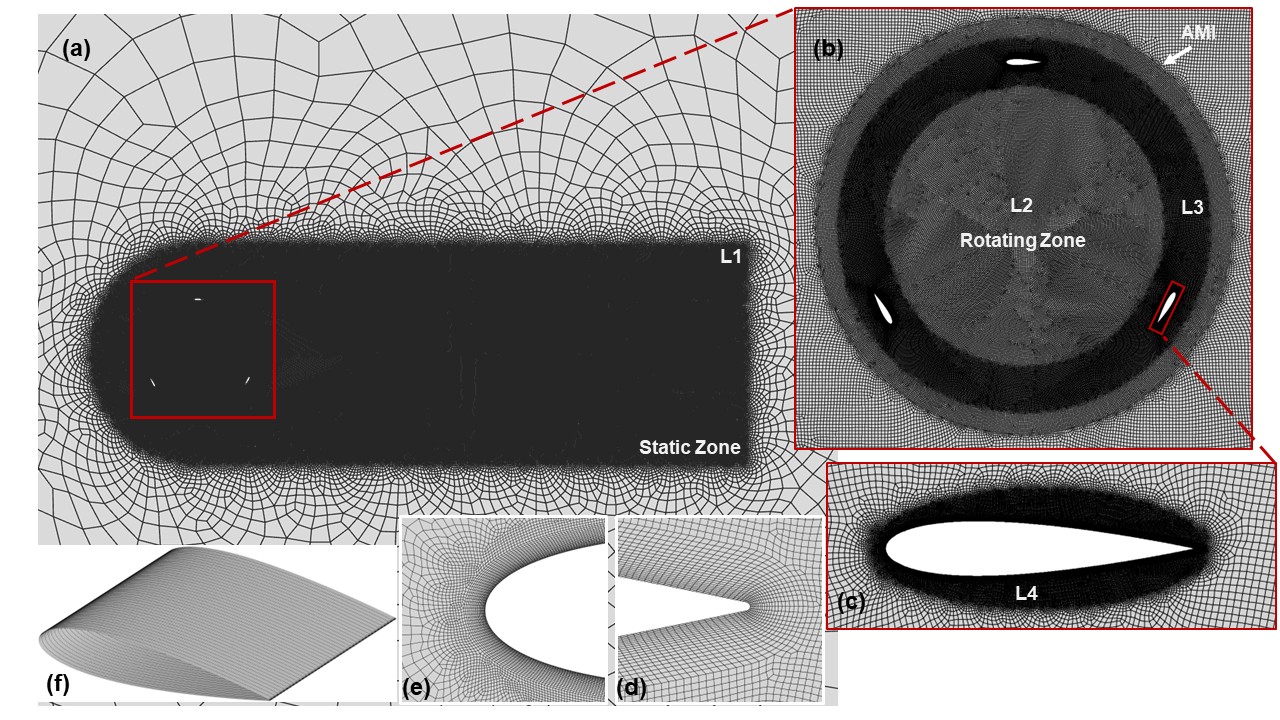}
    \caption{Mesh-refinement regions showing (a) refinement of the wake (L1), (b) refinement of the rotating-zone core (L2) and annular blade-trajectory region (L3), (c) mesh resolution near the blade (L4), (d) trailing-edge refinement, (e) leading-edge refinement, and (f) resolution along the span}
  \label{fig:mesh_rr}
\end{figure}

Grid-convergence tests are carried out on a three-bladed turbine at $U_\infty=8~\mathrm{m/s}$, while $J$ is kept fixed at $0.018~\mathrm{kg\,m^2}$, as reported in the experimental investigation of Rainbird \cite{rainbird2007aerodynamic}. An unstructured, quad-dominant mesh is employed, and a hierarchical refinement strategy is applied to obtain grid-independent solutions. The computational domain is partitioned into a stationary zone and a rotating zone with multiple refinement levels (see Fig.~\ref{fig:mesh_rr}). In the stationary zone, refinement level L1 is applied to the wake region, extending approximately $5D$ downstream of the turbine and $2D$ in the cross-stream direction (Fig.~\ref{fig:mesh_rr}a). Within the rotating zone, nested refinement levels are used to resolve the blade-passage flow. Level L2 refines the core of the rotating zone (Fig.~\ref{fig:mesh_rr}b), level L3 forms an annular refinement band along the trajectory of the blades (Fig.~\ref{fig:mesh_rr}b), and level L4 provides refinement near the surface of the blades (Figs.~\ref{fig:mesh_rr}c - \ref{fig:mesh_rr}f), including enhanced resolution around the leading and trailing edges\edt{,} where strong shear layers form and roll up into coherent vortices.

\edt{For further} near-wall resolution on the surfaces of the blades, inflated quadrilateral layers are introduced. A total of $20$ inflation layers are used. The height of the first layer is selected to maintain $y^+ \approx 1$, with a maximum value of $y^+ \le 5$, and the growth ratio is $1.08$, consistent with established practice \cite{ramirez2021assessment,li20132}.

\begin{figure}
  \centering
  \includegraphics[width=0.8\linewidth]{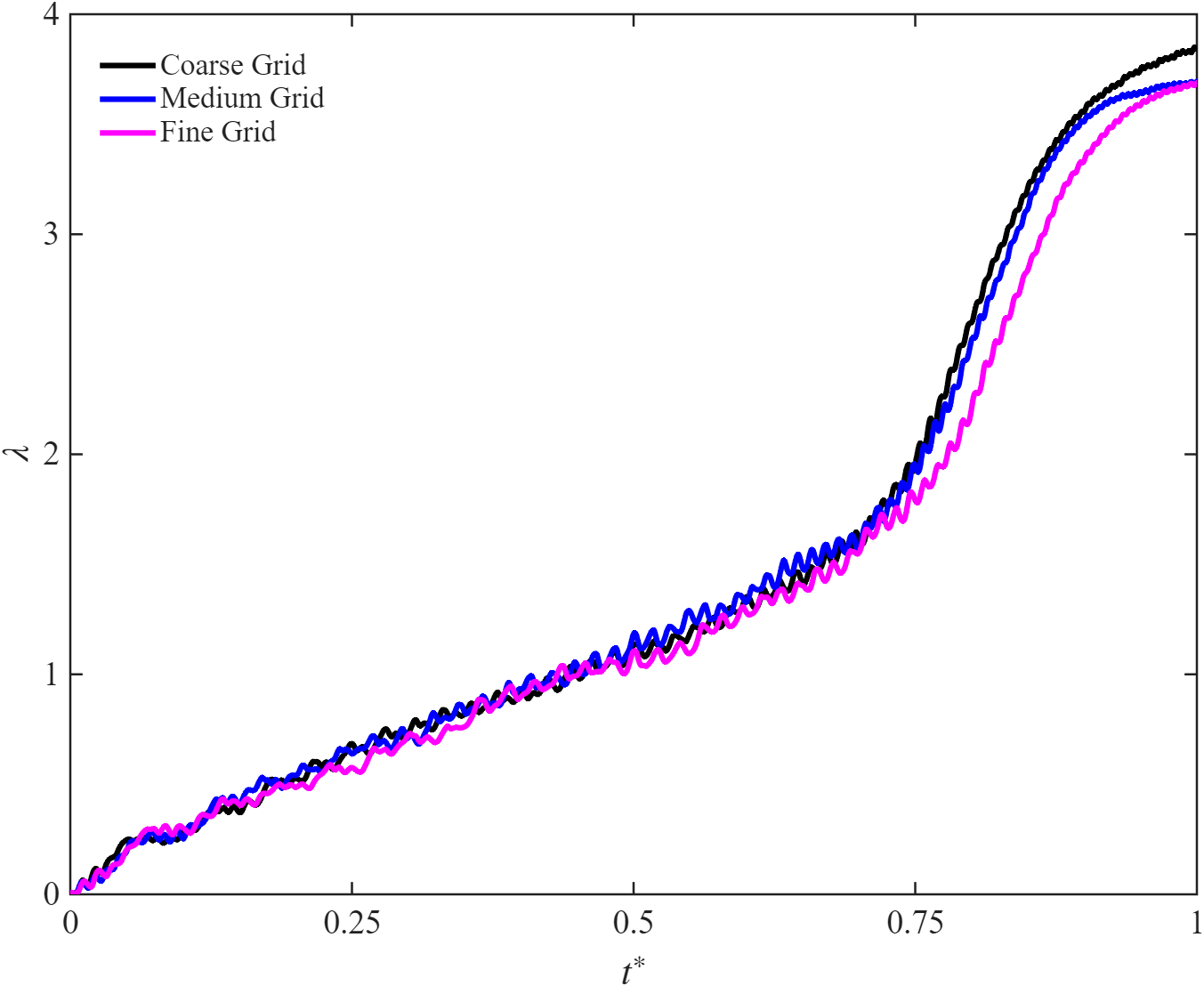}
  \caption{Comparison of \edt{temporal histories} of $\lambda$ \edt{obtained through} the coarse, medium, and fine grids}
  \label{fig:GridInd}
\end{figure}

Three meshes with coarse, medium, and fine resolutions are generated for the grid-convergence study by systematically varying the characteristic cell sizes associated with levels L1 - L4, as summarized in Table~\ref{tab:refinement_levels}. The self-starting response of the \edt{turbine from the} three mesh configurations is compared in Fig.~\ref{fig:GridInd} through the temporal variation of the tip-speed ratio, \edt{defined as} $\lambda=R\omega/U_\infty$. The time axis, presented as $t^{\ast}$, is normalized by the time required for the turbine to reach $\lambda_{steady}$, following previous investigations \edt{on this subject} \cite{asr2016study,khalid2022self}. \edt{For simulating flow-induced rotations of a VAWT}, grid convergence is assessed primarily using the final $\lambda_{steady}$\edt{,} because it represents the converged operating state of the freely rotating turbine \cite{liu2026comprehensive,erkan2025taguchi}. The \edt{temporal} histories of $\lambda$ are also examined to assess consistency in the predicted acceleration response. The medium and fine meshes show close agreement in the final $\lambda_{steady}$ and in the overall acceleration profile, whereas the coarse mesh shows a larger deviation. Therefore, the medium mesh is selected for the time-step independence study and the subsequent simulations\edt{,} because it provides an appropriate balance between numerical accuracy and computational cost.

\begin{table}
\centering
\caption{Characteristic refinement sizes associated with levels L1 - L4 and total cell counts used for the grid-convergence study}
\label{tab:refinement_levels}
\begin{tabular}{lccccc}
\hline
Mesh & L1 (m) & L2 (m) & L3 (m) & L4 (m) & No. of cells \\
     & \(\times 10^{-3}\) & \(\times 10^{-3}\) & \(\times 10^{-3}\) & \(\times 10^{-4}\) & \\
\hline
Coarse & 9.0  & 7.0 & 5.0 & 3.5 & 163{,}000 \\
Medium & 7.5  & 3.5 & 2.0 & 2.5 & 312{,}000 \\
Fine   & 6.0  & 2.0 & 1.5 & 2.0 & 540{,}000 \\
\hline
\end{tabular}
\end{table}

After establishing $\mbox{2D}$ grid convergence, the selected mesh is extruded in the spanwise direction to form the $\mbox{3D}$ computational grids used in the present \edt{work}. The spanwise discretization uses $N_z=40$ divisions, consistent with Li et al. \cite{li20132}, who considered spanwise extents up to $2c$ in their analysis. When the selected medium mesh is extruded with $N_z=40$ spanwise divisions, the resulting $\mbox{3D}$ grid contains approximately $14$ million cells. The height of the turbine is varied using the normalized spanwise extent $H^\ast = H/c$ with $H^\ast \in \{1,\,1.5,\,2\}$. \edt{It} preserves the converged in-plane resolution while enabling a controlled assessment of spanwise flow development and vortex evolution during start-up.

\begin{figure}
  \centering
  \includegraphics[width=0.8\linewidth]{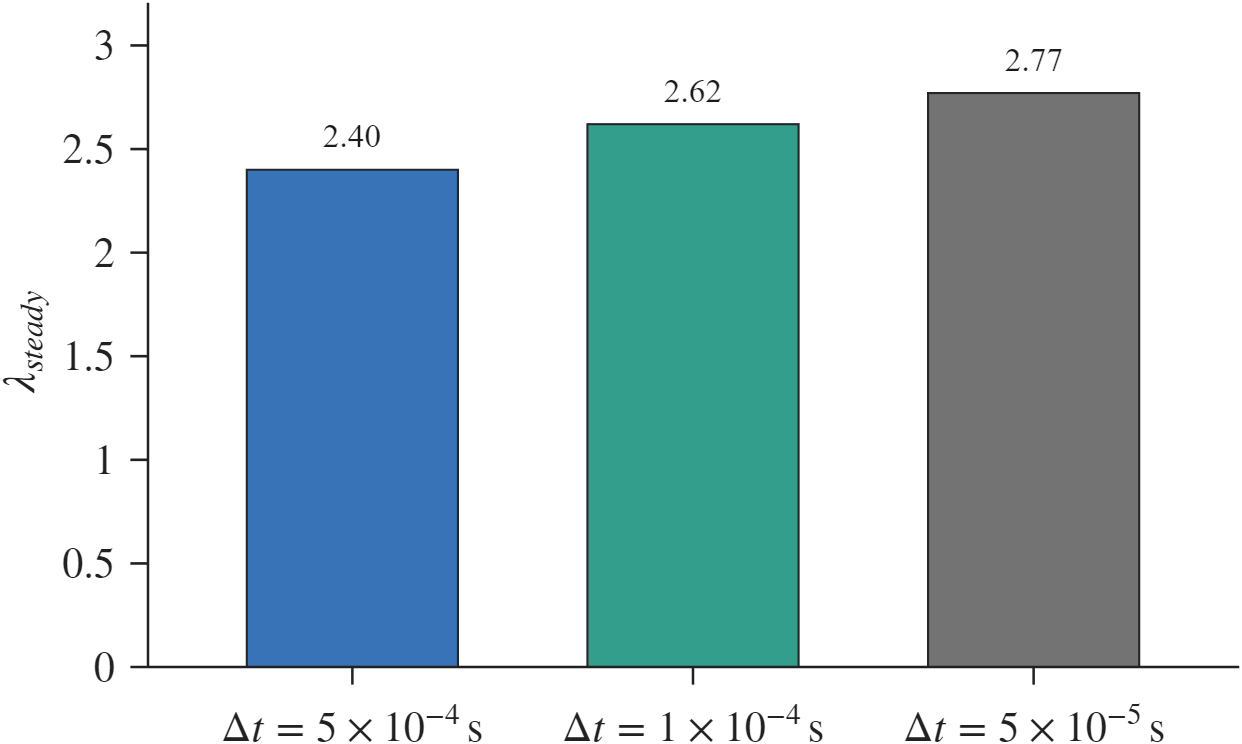}
  \caption{Time-step sensitivity of $\lambda_{steady}$ obtained using \(\Delta t = 5\times10^{-4}\,\mathrm{s}\), \(1\times10^{-4}\,\mathrm{s}\), and \(5\times10^{-5}\,\mathrm{s}\)}
  \label{fig:TimeInd}
\end{figure}

A time-step independence study is performed using the selected medium mesh. In \edt{simulations for} flow-induced self-starting, the time-step size influences the predicted acceleration history because $\omega$ of the rotor is obtained through the time integration of aerodynamic moment rather than being prescribed. \edt{Therefore,} Changes in temporal resolution can alter both the transient \edt{variations in} $\lambda$ and the final $\lambda_{steady}$, as also reported in previous flow-induced \edt{rotation of VAWTs related} studies \cite{liu2026comprehensive,erkan2025taguchi}. \edt{Here, we consider} three time-step sizes, \edt{including} $\Delta t_1=0.0005~\mathrm{s}$, $\Delta t_2=0.0001~\mathrm{s}$, and $\Delta t_3=0.00005~\mathrm{s}$. The resulting values of $\lambda_{steady}$ are $2.40$, $2.62$, and $2.77$, respectively, as shown in Fig.~\ref{fig:TimeInd}. The difference between the two smallest time steps is approximately $5.7\%$, which is comparable to the time-step sensitivity reported by Erkan et al. \cite{erkan2025taguchi}. The case with $\Delta t_2=0.0001~\mathrm{s}$ reproduces the overall acceleration response and predicts a final $\lambda_{steady}$ \edt{effectively} while substantially reducing the computational cost of the $\mbox{3D}$ $\mbox{LES}$. Therefore, $\Delta t_2=0.0001~\mathrm{s}$ is selected for the simulations in the present study as a balance between temporal resolution and computational cost.

\begin{figure}
  \centering
  \includegraphics[width=0.9\linewidth]{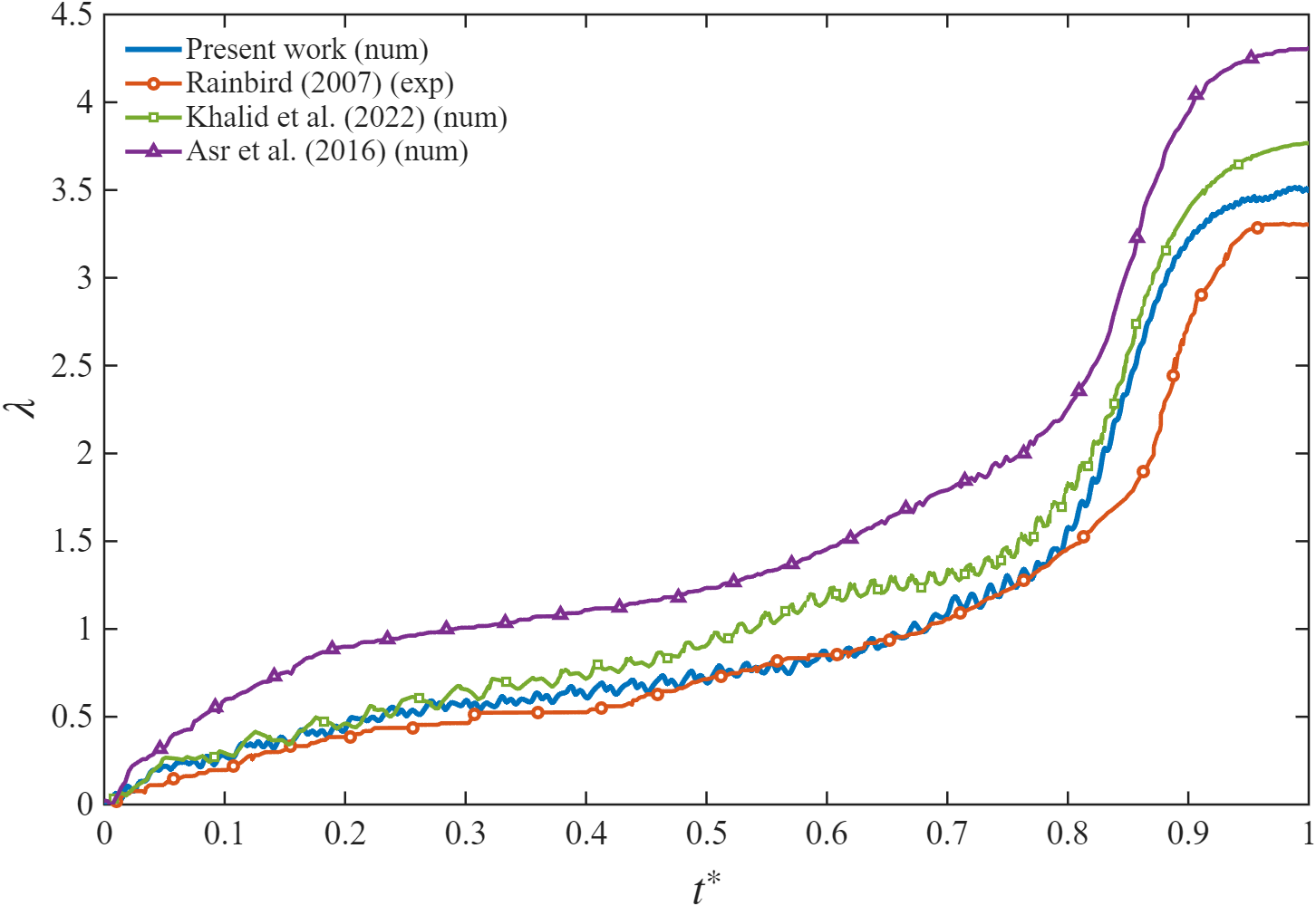}
  \caption{Comparison of the predicted $\lambda$ with the experimental and numerical results reported in the literature}
  \label{fig:val}
\end{figure}

The present computational framework is validated \edt{through} the experimental work of Rainbird \cite{rainbird2007aerodynamic} and the numerical studies conducted by Khalid et al. \cite{khalid2022self} and Asr et al. \cite{asr2016study}. \edt{For this purpose,} the geometry of \edt{the VAWT} consists of three NACA0018 \edt{blades} subjected to $U_\infty=6~\mathrm{m/s}$ with $c=83~\mathrm{mm}$, $R=0.375~\mathrm{m}$, and $J=0.018~\mathrm{kg\,m^2}$. As shown in Fig.~\ref{fig:val}, the \edt{four} established self-starting stages discussed by Du et al. \cite{du2019review} are captured, including an initial acceleration stage, a transient plateau (dead-band) with only a slow increase in $\lambda$, a subsequent rapid-acceleration stage, and finally convergence to a quasi-steady operating value of $\lambda$.

Overall, the present results show close agreements with the experimental\edt{ly determined profiles of $\lambda$} and remain consistent with \edt{the} previously reported simulations. It is noted that the experiment \edt{involved} a rotor with a finite span and includes resistive torques associated with the test rig and drivetrain, whereas the $\mbox{2D}$ \edt{simulations} presented here assume an infinite span and a freely rotating turbine without structural or generator damping. These idealizations can lead to mild overprediction relative to the experiments. However, the close correspondence with Rainbird \cite{rainbird2007aerodynamic} and the agreement with Khalid et al. \cite{khalid2022self} and Asr et al. \cite{asr2016study} provide confidence in the present computational setup and support its use for the subsequent analysis of vortex dynamics during self-starting.

\section{Results and Discussion}
\label{sec:results}

Before presenting and \edt{explaining our} results, the behavior of the flow around \edt{a} VAWT during self-starting is first compared with findings reported in the literature to establish a firm basis for the subsequent discussion. The response of the \edt{flow} during self-starting is then examined at two scales. The overall evolution of vortices within the rotating region and the development of the wake are first analyzed. The analysis is subsequently focused on the flow near a single blade by examining the evolution of vortical structures during the previously identified regimes \edt{in the} self-starting \edt{process}.

\begin{figure}[h!]
  \centering
  \includegraphics[width=0.45\linewidth]{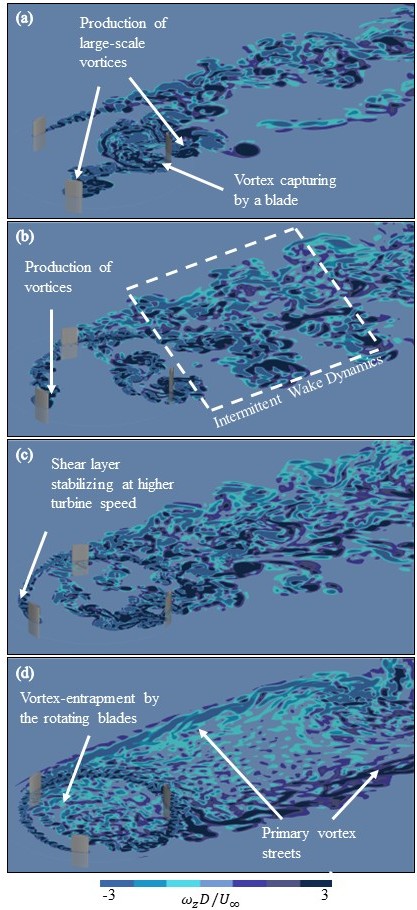}
    \caption{Development of vortical activity during self-starting, consistent with the results reported by Khalid et al. \cite{khalid2022self}. The mid-plane shows the dimensionless spanwise vorticity $\omega_z D/U_\infty$ for $H^{*}=2$}
  \label{fig:comp}
\end{figure}

To analyze the overall vortex dynamics and reinforce confidence in the numerical setup, the evolution and convection of vortices are compared with results reported in the literature. The \edt{2D numerical investigations for flow-induced rotations} of single- and dual-stage turbines by Khalid et al. \cite{khalid2022self} are selected for this comparison. To the best of the authors’ knowledge, this is the only study that discusses the production and decay of vortices in a self-starting VAWT together with the development of the established wake under steady conditions. They reported the production of large-scale vortices at the onset of rotation and an intermittent state of the wake. As the turbine rotates at higher speeds, the wake of a single blade begins to stabilize and the shedding of vortices is reduced. Once the turbine achieves self-starting characteristics, the wake settles into two vortex streets formed \edt{on} the windward and leeward \edt{sides}. To examine these patterns, a mid-span section of the \edt{VAWT with} $H^{*}=2$ is analyzed using contours of vorticity, as shown in Fig.~\ref{fig:comp}. The initial acceleration stage is characterized by the production of large-scale vortices (Fig.~\ref{fig:comp}a), followed by an intermittent wake during the dead-band (Fig.~\ref{fig:comp}b). As the turbine accelerates, the shear layers shed by a single blade begin to stabilize during the rapid-acceleration stage (Fig.~\ref{fig:comp}c). Finally, the entrapment of vortices by the rotating blades is observed as the wake settles into two vortex streets during \edt{the} steady \edt{rotation} (Fig.~\ref{fig:comp}d). This comparison shows that the response of the fluid to the rotation of the turbine \edt{in the prescribed manner, as introduced here} is consistent with that reported by Khalid et al. \cite{khalid2022self} \edt{through their simulations for flow-induced rotations of VAWTs}.

The \edt{temporal} histories of $\lambda$ and the effective angle of attack ($\alpha_{eff}$) are next examined to determine whether the prescribed variation in $\omega$ of the rotor reproduces the main stages observed for a similar turbine undergoing flow-induced conditions. The aerodynamic behavior of a blade is strongly influenced by the temporal variation of $\alpha_{eff}$, which is expressed as:

\begin{equation}
\alpha_{eff}=\arctan(\frac{\sin{\theta}}{\cos{\theta+\lambda})})
\label{eq:alpha_eff}
\end{equation}

\begin{figure}[h!]
  \centering
  \includegraphics[width=1\linewidth]{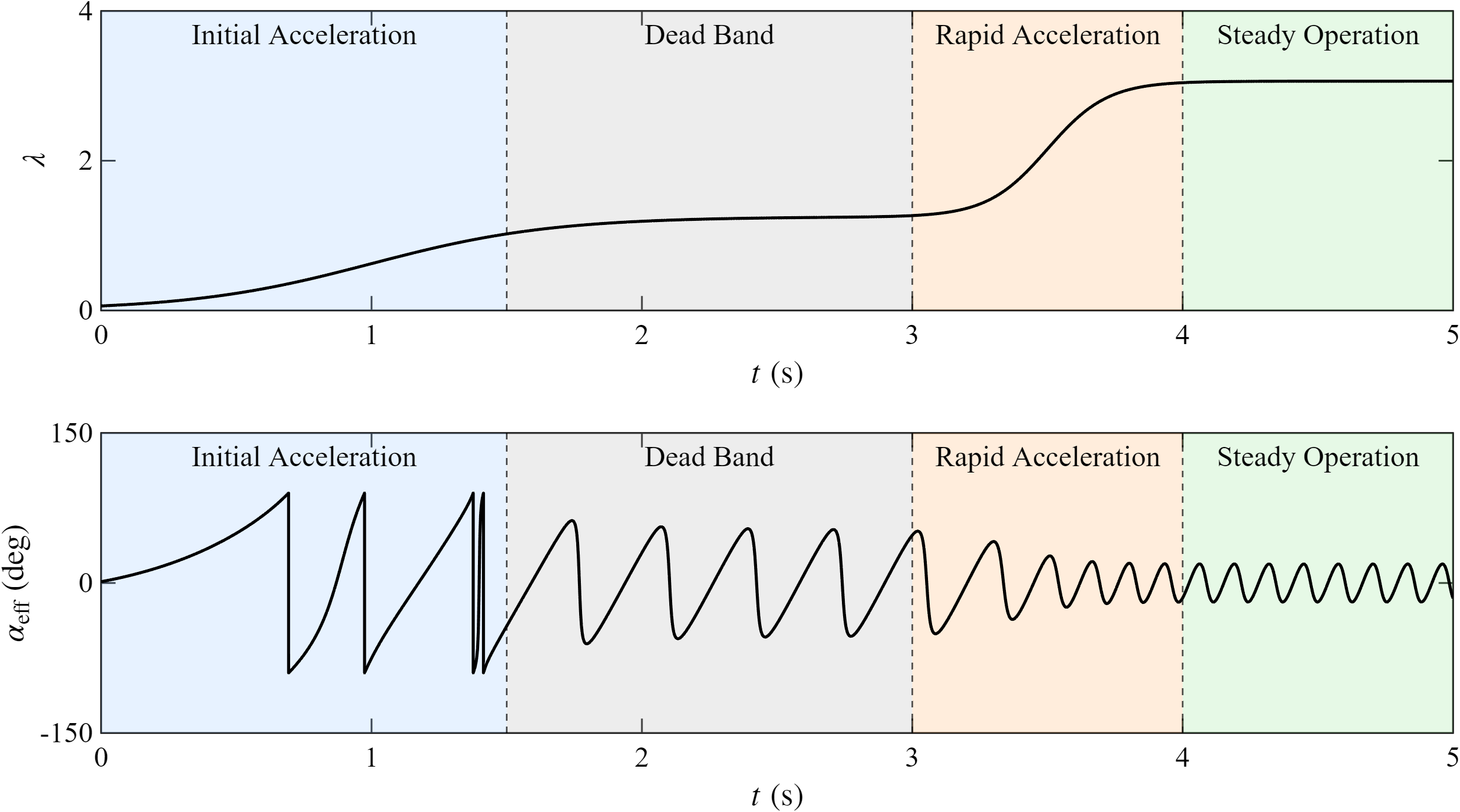}
  \caption{Time histories of $\lambda$ and $\alpha_{eff}$ for the present work, in which the angular velocity of the rotor is prescribed, with the principal stages of self-starting identified}
  \label{fig:alphaeffts_c}
\end{figure}

\begin{figure}[h!]
  \centering
  \includegraphics[width=1\linewidth]{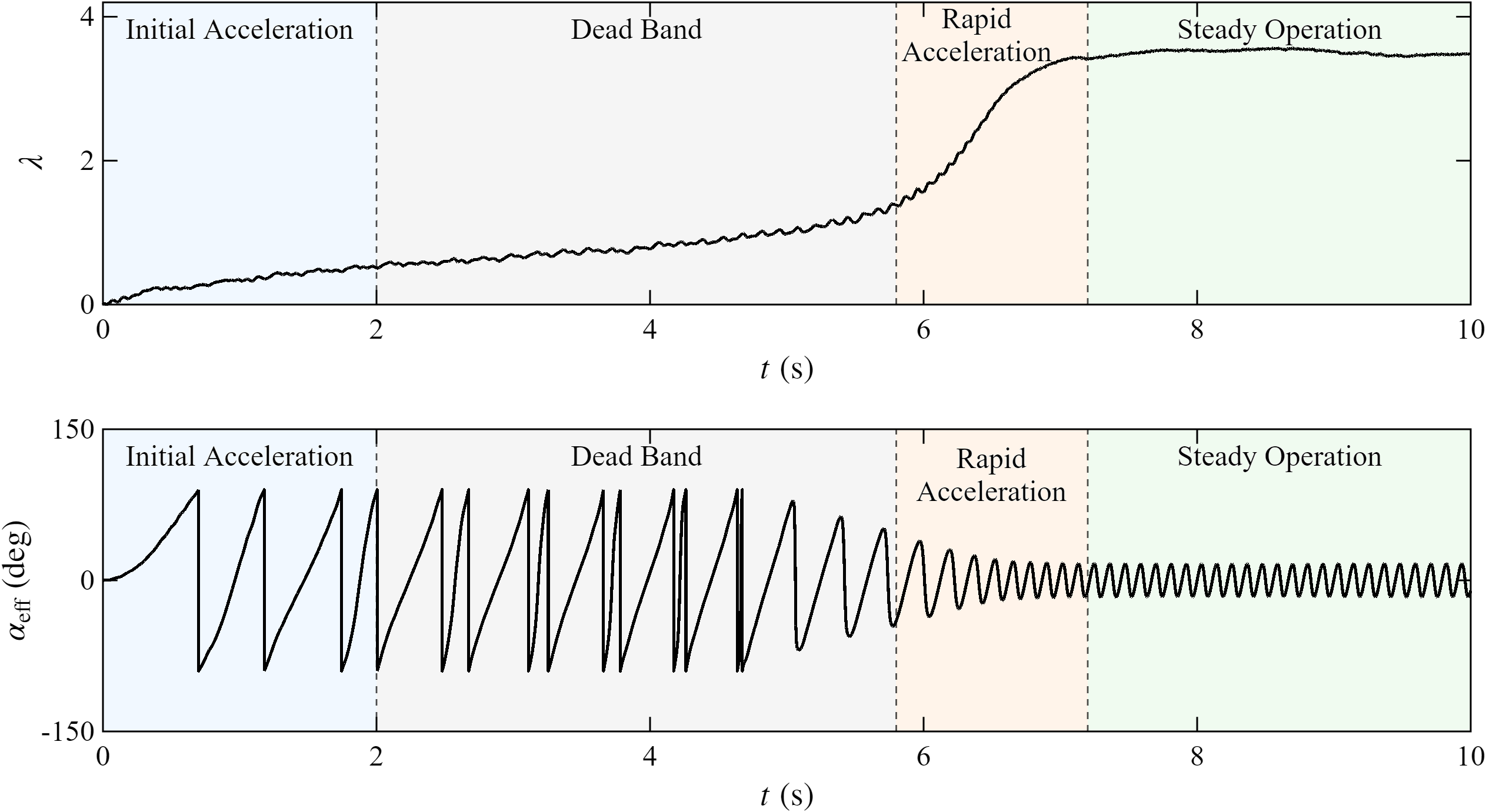}
  \caption{Time histories of $\lambda$ and $\alpha_{eff}$ for a self-starting $\mbox{2D}$ VAWT under flow-induced conditions with $c=83\,\mathrm{mm}$ and $U_{\infty}=6\,\mathrm{m/s}$}
  \label{fig:alphaeffts_s}
\end{figure}

When $\alpha_{eff}$ exceeds the static stall angle ($\alpha_{\mathrm{ss}}$), flow reversal begins within the boundary layer near the trailing edge on the suction side of the blade, \edt{which indicates} the onset of dynamic stall \cite{leishman2006principles}. For NACA0018 airfoil, $\alpha_{\mathrm{ss}}$ \edt{was} reported to range from approximately $13^{\circ}$ to $20^{\circ}$, depending on the Reynolds number \cite{timmer2008two,stangfeld2015unsteady,damiola2023influence,le2021dynamics}. The time histories of $\lambda$ and $\alpha_{eff}$ obtained in the present study, in which $\omega$ of the rotor is prescribed, are presented in Fig.~\ref{fig:alphaeffts_c}. The corresponding histories for a 2D turbine with the same geometry and inlet velocity under flow-induced conditions \cite{muhammad2026geometric} are presented in Fig.~\ref{fig:alphaeffts_s}. The purpose of this comparison is not to reproduce the exact duration of self-starting. Instead, it is used to determine whether the prescribed variation in $\omega$ reproduces the sequence of stages and the corresponding variation of $\alpha_{eff}$ observed under flow-induced conditions. The duration of the process is reduced in the present study to lower the computational cost while providing sufficient time for the vortices to evolve and convect during the principal stages of self-starting.

During initial acceleration, $\lambda$ increases towards unity while $\alpha_{eff}$ undergoes large variations because $\omega$ of the rotor remains relatively low. The positive values of $\alpha_{eff}$ correspond to the upwind half of the \edt{rotational} cycle from $\theta=0^\circ$ to $180^\circ$, while the negative values correspond to the downwind half of $\theta=180^\circ$ to $360^\circ$. During the dead-band, $\lambda$ changes only slightly and the turbine rotates at an approximately constant $\omega$. The frequency of the variation in $\alpha_{eff}$ increases at this stage, although its amplitude remains large enough for $\alpha_{eff}$ to exceed $\alpha_{\mathrm{ss}}$ over a considerable part of the cycle. As the turbine approaches \edt{its} rapid acceleration \edt{phase}, $\lambda$ increases sharply and the range of $\alpha_{eff}$ progressively narrows \edt{down}. Its variation approaches a small-amplitude sinusoidal form, which reduces the portion of the cycle over which $\alpha_{eff}$ exceeds $\alpha_{\mathrm{ss}}$ and consequently reduces the tendency for extensive flow separation. During the quasi-steady operation, $\lambda$ approaches an approximately constant value and $\alpha_{eff}$ exhibits repeatable periodic variations of relatively small amplitude. The same overall progression is observed in the flow-induced case, which supports the use of the \edt{newly proposed} prescribed variation in $\omega$ of the rotor to represent the principal stages of self-starting.

The analysis now proceeds to examine the flow behavior and the vortical structures for the three spanwise extents \edt{of the blades in a VAWT} considered in this study. The initial acceleration, dead-band, rapid acceleration, and quasi-steady operating regimes are discussed in Sections~\ref{subsec:res_accel}, \ref{subsec:res_deadband}, \ref{subsec:res_rapid}, and \ref{subsec:res_steady}, respectively. During each stage, a single blade is tracked from the reference position $\theta=0^\circ$, and the same analysis is performed for all three spanwise extents. The vortical structures are examined at $\theta=60^\circ$, $120^\circ$, $180^\circ$, and $240^\circ$ because these azimuthal positions are associated with pronounced flow separation and dynamic stall activity in VAWTs \cite{ferreira2007simulating,le2022dynamic}. This analysis enables a direct comparison of the effects of the spanwise extent on the formation, coherence, detachment, convection, and breakdown of the vortical structures throughout the principal stages of self-starting.

Before discussing the evolution of the vortical structures near the blade, the terminology used for the two surfaces of the blade and the sign convention for vorticity are clarified \edt{here}. The chord line separates the blade into an inner surface facing the shaft of the turbine and an outer surface facing away from it. Under the sign convention adopted in the vorticity contours, positive vorticity develops along the inner surface at the beginning of the rotation cycle at $\theta=0^\circ$, whereas negative vorticity develops along the outer surface. The aerodynamic roles of the two surfaces vary continuously\edt{, in terms of being the suction and pressure sides,} as the blade moves through a complete revolution. During the upwind half-cycle ($0^\circ \leq \theta < 180^\circ$), the inner surface acts predominantly as the suction side, while the outer surface acts as the pressure side. During the downwind half-cycle ($180^\circ < \theta \leq 360^\circ$), these aerodynamic roles are reversed. At $\theta=180^\circ$, the blade is at the transition between the two half-cycles, and\edt{, therefore,} the inner and outer surfaces are in the process of interchanging their aerodynamic roles rather than being distinctly established as either the suction or pressure side.

\subsection{Initial Acceleration Regime}
\label{subsec:res_accel}

The initial acceleration regime represents the beginning of the self-starting process, during which the rotor gains angular speed\edt{,} and the wake and vortical structures near the blade begin to develop. Since $\omega$ remains relatively low during this regime, the blade experiences large variations in $\alpha_{eff}$ over each rotation cycle. Vortex dynamics is first examined for $H^{*}=1$, followed by $H^{*}=1.5$ and $H^{*}=2$, to determine how increasing the span \edt{of a blade} influences the development, coherence, convection, and breakdown of the vortical structures.

\begin{figure}[h!]
  \centering
  \includegraphics[width=1\linewidth]{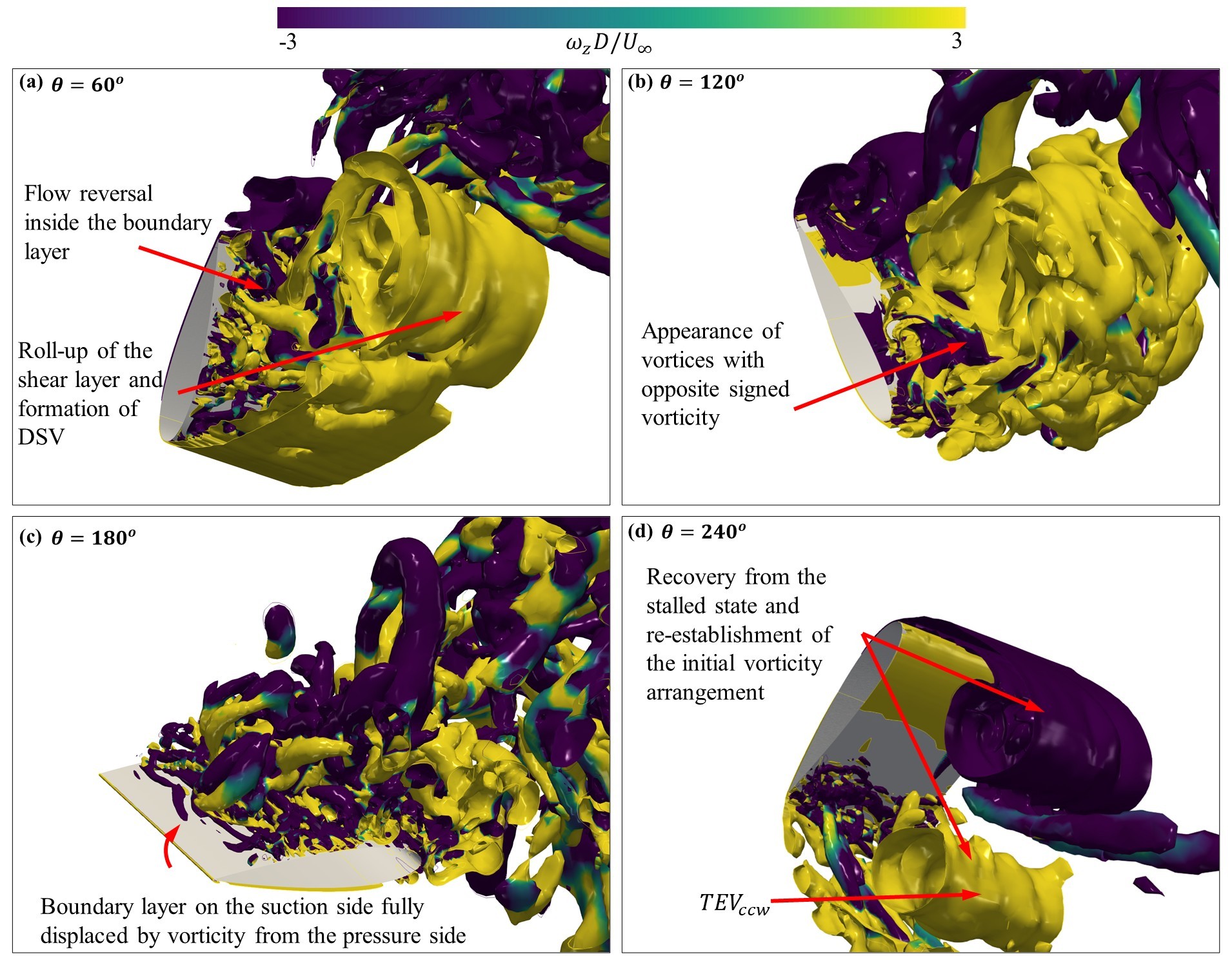}
  \caption{Iso-surfaces of the $Q$-criterion at $Q=10{,}000$ showing the \edt{flow} field during the initial acceleration regime, colored by dimensionless spanwise vorticity $\omega_z D/U_\infty$ for $H^{*}=1$}
  \label{fig:initAcc_1c}
\end{figure}

For $H^{*}=1$, flow reversal begins within the boundary layer near the trailing edge on the suction side of the blade at $\theta=60^\circ$\edt{, as exhibited} in Fig.~\ref{fig:initAcc_1c}a, where $\alpha_{eff}$ is considerably greater than $\alpha_{\mathrm{ss}}$. The separated shear layer rolls up to form a dynamic stall vortex (DSV), which remains attached to the blade and continues to grow as the blade advances through the upwind half-cycle. At $\theta=120^\circ$ in Fig.~\ref{fig:initAcc_1c}b, the DSV \edt{increases} considerably in size while remaining largely coherent. \edt{Simultaneously}, secondary vortices with opposite-signed vorticity develop between the DSV and the blade. Their interaction with the DSV limits its continued growth and promotes its detachment from the blade and subsequent convection into the wake, as also noted by Le Fouest and Mulleners \cite{le2022dynamic}. At $\theta=180^\circ$ in Fig.~\ref{fig:initAcc_1c}c, the boundary layer associated with the preceding upwind half-cycle is displaced as \edt{the} fluid with vorticity of the opposite sign advances around the trailing edge. The flow at this position consequently undergoes a reorganization of the boundary layers as the aerodynamic roles of the inner and outer surfaces interchange. As the blade advances to $\theta=240^\circ$ in Fig.~\ref{fig:initAcc_1c}d, the aerodynamic roles associated with the downwind half-cycle become more clearly established, with the inner surface acting as the pressure side and the outer surface as the suction side. \fmrev{Within this transition, the flow over the inner half of the blade reattaches earlier, between $\theta=180^\circ$ and $240^\circ$, and this reattachment is accompanied by the shedding of a counterclockwise trailing-edge roller vortex ($TEV_{ccw}$), as shown in Fig.~\ref{fig:initAcc_1c}d. At the same time, the separated shear layer near the leading edge rolls up and begins to reattach over the outer half of the blade. As $\alpha_{\mathrm{eff}}$ decreases below $\alpha_{ss}$ later in the cycle beyond $\theta=240^\circ$, the flow over the outer half gets attached to the surface of the blade. This reattachment is followed by the shedding of a clockwise leading-edge roller vortex ($LEV_{cw}$), which occurs beyond the azimuthal position shown in Fig.~\ref{fig:initAcc_1c}d. Consequently, the distribution of vorticity begins to reorganize toward its initial arrangement as the blade approaches the completion of the rotation cycle, with negative vorticity developing along the outer surface and positive vorticity developing along the inner surface of the blade.}

\begin{figure}[h!]
  \centering
  \includegraphics[width=1\linewidth]{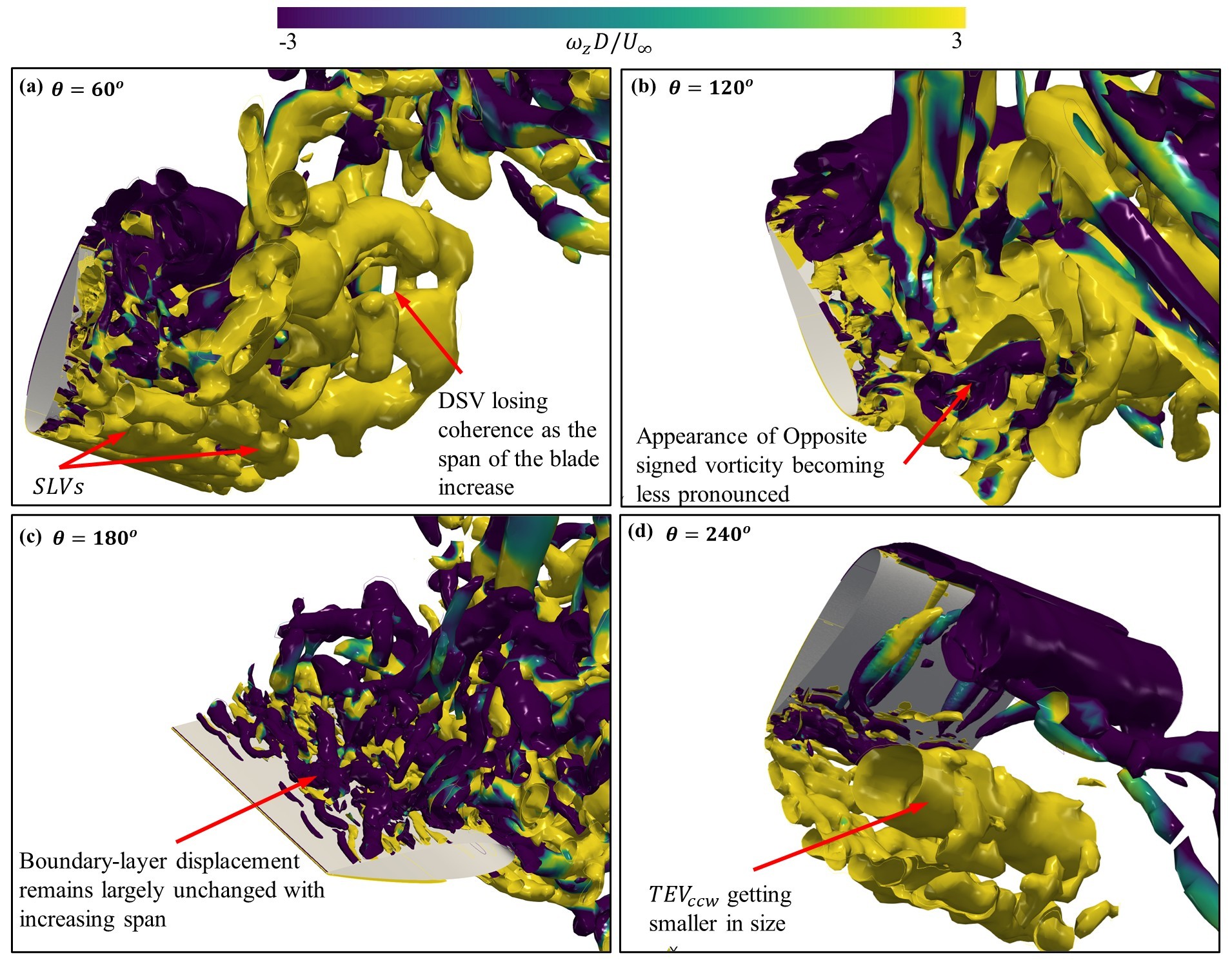}
  \caption{Iso-surfaces of the $Q$-criterion at $Q=10{,}000$ showing the \edt{the flow} field during the initial acceleration regime, colored by dimensionless spanwise vorticity $\omega_z D/U_\infty$ for $H^{*}=1.5$}
  \label{fig:initAcc_1.5c}
\end{figure}

\edt{Increasing the span of the blade to} $H^{*}=1.5$, the same general sequence of flow development is observed, although \edt{it} modifies the organization of the vortical structures. At $\theta=60^\circ$ in Fig.~\ref{fig:initAcc_1.5c}a, the DSV exhibits a noticeable loss of coherence along the span compared with \edt{the case of} $H^{*}=1$. The shear layer associated with the formation of the DSV \edt{experiences} greater variation\edt{s} along the span and begins to \edt{roll quickly to} shed \fm{multiple \edt{vortices, identified here as $SLVs$} near the leading edge,} which subsequently convect away from the blade. At $\theta=120^\circ$ in Fig.~\ref{fig:initAcc_1.5c}b, the vortices with opposite-signed vorticity that develop between the DSV and the surface of the blade become less distinct, while the DSV exhibits greater variation along the span. At $\theta=180^\circ$ in Fig.~\ref{fig:initAcc_1.5c}c, stronger \edt{vortex-vortex interactions are observed} along the span between regions of fluid with \edt{oppositely-signed vorticity} as the aerodynamic roles of the inner and outer surfaces \edt{are interchanged}. At $\theta=240^\circ$ in Fig.~\ref{fig:initAcc_1.5c}d, \fm{the flow \edt{is} already reattached to the inner surface of the blade, resulting in the shedding of a $TEV_{ccw}$, as observed for $H^*=1$. Over the outer half of the blade,} the flow begins to reattach as $\alpha_{eff}$ decreases below $\alpha_{\mathrm{ss}}$, and sheds a $LEV_{cw}$  later in the cycle. \edt{A particular feature of \fm{the $TEV_{ccw}$} is its smaller size compared to the one shed from the blade with $H^{*}=1$ (see Fig.~\ref{fig:initAcc_1c}\fm{d}).}

\begin{figure}[h!]
  \centering
  \includegraphics[width=1\linewidth]{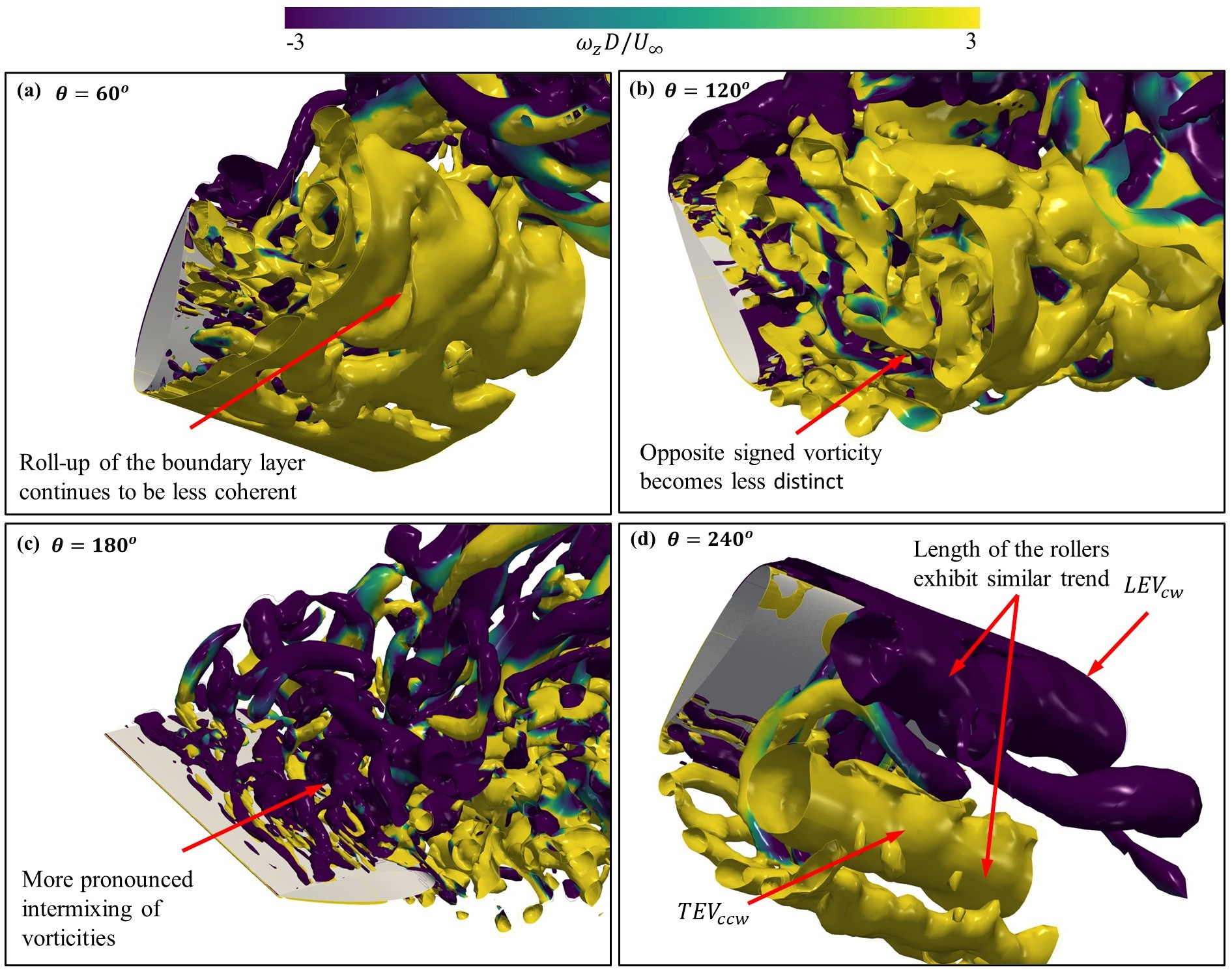}
  \caption{Iso-surfaces of the $Q$-criterion at $Q=10{,}000$ showing the \edt{flow} field during the initial acceleration regime, colored by dimensionless spanwise vorticity $\omega_z D/U_\infty$ for $H^{*}=2$}
  \label{fig:initAcc_2c}
\end{figure}

For \edt{a span} $H^{*}=2$ \edt{for the blade}, the influence of the increased spanwise extent becomes more pronounced. At $\theta=60^\circ$ in Fig.~\ref{fig:initAcc_2c}a, the DSV becomes less organized than for the smaller spanwise extents, indicating that stronger instabilities develop along the span. At $\theta=120^\circ$ in Fig.~\ref{fig:initAcc_2c}b, the secondary vortices with \edt{oppositely-signed} vorticity that develop between the DSV and the surface of the blade become less pronounced, while the DSV exhibits reduced coherence along the span. At $\theta=180^\circ$ in Fig.~\ref{fig:initAcc_2c}c, more pronounced vortex braids develop as the boundary layers along the inner and outer surfaces intermix while their aerodynamic roles interchange. At $\theta=240^\circ$ in Fig.~\ref{fig:initAcc_2c}d, \fm{the same sequence of shear-layer roll-up and flow reattachment is observed, leading to the shedding of $TEV_{ccw}$. \edt{The flow structure} $TEV_{ccw}$ has a size comparable to that observed for $H^*=1.5$, but \edt{it} exhibits \edt{more significant} undulations along the span, which \edt{may be} attributed to \edt{development of stronger} three-dimensional instabilities \cite{gungor2024effect,michelis2018origin}.}

\subsection{Dead-Band (Plateau) Regime}
\label{subsec:res_deadband}

Following the initial acceleration regime, the rotor enters the dead-band, during which $\omega$ remains nearly constant and increases only weakly with time. Although the blade continues to experience large variations in $\alpha_{eff}$, the vortex dynamics differs considerably from that observed during initial acceleration. \edt{Particularly}, the roll-up of the separated shear layer becomes less pronounced, the development of the DSV is suppressed, and stronger \edt{interactions} occur between the blade and vortices generated earlier in the rotation cycle. These {blade-vortex interactions (BVI)} maintain a highly disturbed flow around the blade and are associated with the limited increase in $\omega$ during this regime. The evolution of the vortical structures is first examined for $H^{*}=1$, followed by $H^{*}=1.5$ and $H^{*}=2$, to determine how it changes with \edt{an} increasing \edt{span of the blade}.

\begin{figure}[h!]
  \centering
  \includegraphics[width=1\linewidth]{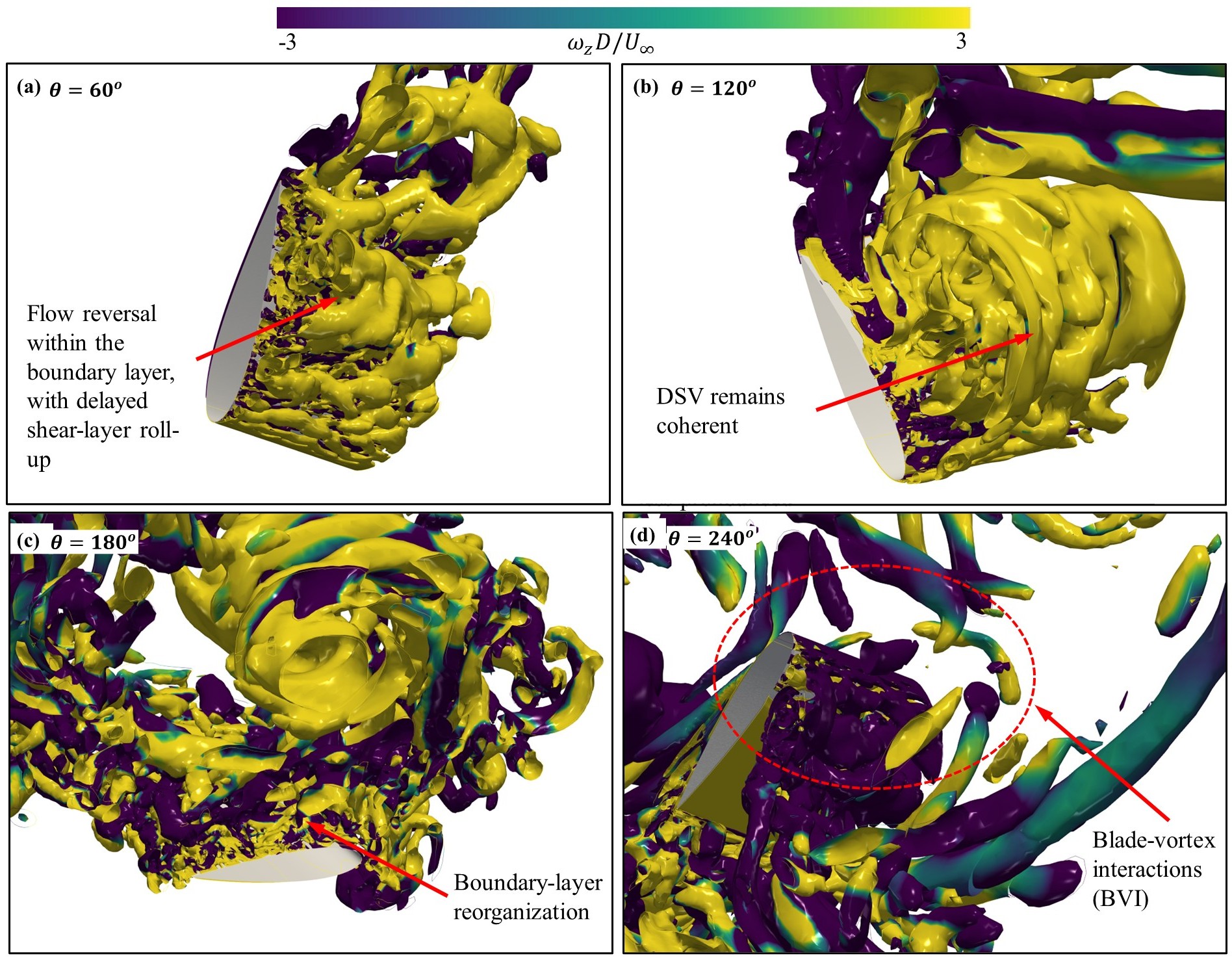}
  \caption{Iso-surfaces of the $Q$-criterion at $Q=10{,}000$ showing the vortex field during the dead-band regime, colored by dimensionless spanwise vorticity $\omega_z D/U_\infty$ for $H^{*}=1$}
  \label{fig:deadBand_1c}
\end{figure}

For \edt{the blade with} $H^{*}=1$, flow reversal is evident within the boundary layer near the trailing edge at $\theta=60^\circ$ in Fig.~\ref{fig:deadBand_1c}a. However, the separated shear layer does not undergo the pronounced roll-up observed during the initial acceleration regime. At $\theta=120^\circ$ in Fig.~\ref{fig:deadBand_1c}b, the DSV remains close to the surface of the blade and retains considerable coherence, although its growth is suppressed and its subsequent detachment is delayed owing to the higher $\omega$ of the rotor compared with that during initial acceleration. At $\theta=180^\circ$ in Fig.~\ref{fig:deadBand_1c}c, the boundary layers reorganize as the aerodynamic roles of the inner and outer surfaces interchange during the transition between the upwind and downwind half-cycles. At $\theta=240^\circ$ in Fig.~\ref{fig:deadBand_1c}d, vortices \fm{that} \edt{are} shed earlier in the rotation cycle interact with the blade, disturbing the flow near its surface and hindering reattachment.

\begin{figure}[h!]
  \centering
  \includegraphics[width=1\linewidth]{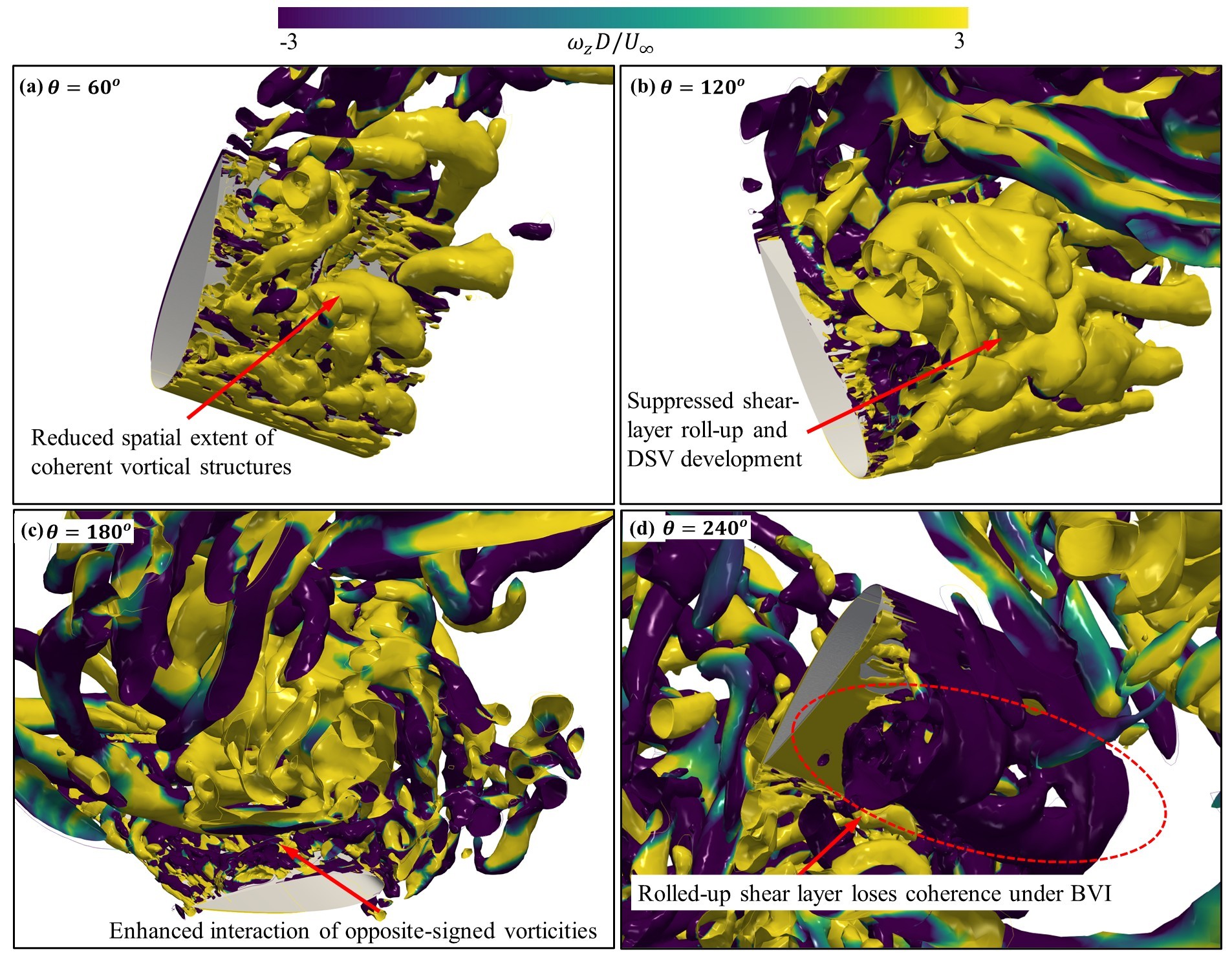}
  \caption{Iso-surfaces of the $Q$-criterion at $Q=10{,}000$ showing the vortex field during the dead-band regime, colored by dimensionless spanwise vorticity $\omega_z D/U_\infty$ for $H^{*}=1.5$ \fm{figure changed (14d)}}
  \label{fig:deadBand_1.5c}
\end{figure}

For \edt{the case with} $H^{*}=1.5$, the same general sequence of \edt{events} is observed, although the increase in the spanwise extent modifies the organization of the vortical structures. At $\theta=60^\circ$ in Fig.~\ref{fig:deadBand_1.5c}a, flow reversal remains evident near the trailing edge, while the roll-up of the separated shear layer remains suppressed and the vortical structures exhibit reduced coherence and greater deformation along the span compared with \edt{the previously explained case}. At $\theta=120^\circ$ in Fig.~\ref{fig:deadBand_1.5c}b, the DSV remains close to the surface of the blade but becomes less coherent along the span, with stronger interaction between the DSV, opposite-signed vorticity, and the surrounding \edt{secondary} vortical structures. At $\theta=180^\circ$ in Fig.~\ref{fig:deadBand_1.5c}c, the blade is in the transitional aerodynamic state between the upwind and downwind half-cycles, and enhanced mixing of opposite-signed vorticities is observed as the roles of the inner and outer surfaces interchange. At $\theta=240^\circ$ in Fig.~\ref{fig:deadBand_1.5c}d, vortices \edt{get} shed during the preceding upwind half-cycle convect downstream and subsequently interact again with the blade. \fm{\edt{We observe that} BVI is more intense \edt{here} than that observed for the \edt{case with a} smaller span and disrupts the roll-up of the separated shear layer, which is no longer coherent along the span.}

\begin{figure}[h!]
  \centering
  \includegraphics[width=1\linewidth]{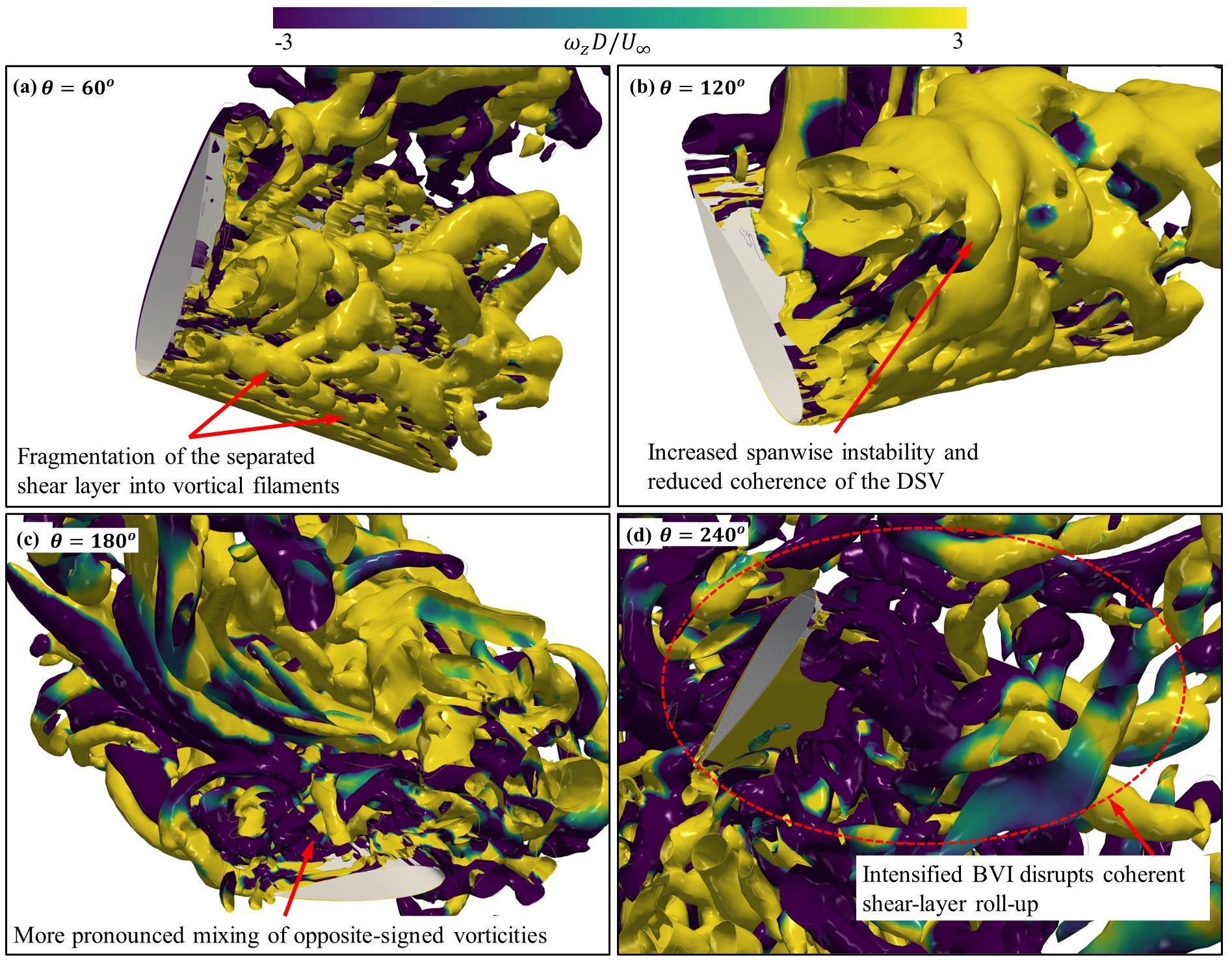}
  \caption{Iso-surfaces of the $Q$-criterion at $Q=10{,}000$ showing the vortex field during the dead-band regime, colored by dimensionless spanwise vorticity $\omega_z D/U_\infty$ for $H^{*}=2$}
  \label{fig:deadBand_2c}
\end{figure}

\fm{Increasing the \edt{blade's span} to $H^{*}=2$ does not alter the overall \edt{sequence of flow-related processes in the} dead-band \edt{phase}. \edt{However, we describe some} noticeably changes \edt{in} the coherence and interaction\edt{s} of the vortical structures.} At $\theta=60^\circ$ in Fig.~\ref{fig:deadBand_2c}a, \fm{the separated shear layer becomes more fragmented than \edt{those} in the two \edt{others cases with smaller spans of the blades. Here, the shear layer gets}, broken into smaller vortical filaments and showing a further loss of coherence along the span.} At $\theta=120^\circ$ in Fig.~\ref{fig:deadBand_2c}b, \fm{the DSV exhibits a further loss of coherence and stronger deformation along the span, while the surrounding vortical structures become less organized.} At $\theta=180^\circ$ in Fig.~\ref{fig:deadBand_2c}c, the blade again passes through the transitional aerodynamic state between the two half-cycles, with more pronounced \edt{interferance} of \edt{the oppositely}-signed vorticities\edt{,} as compared \edt{to those} with $H^{*}=1$ and $H^{*}=1.5$. At $\theta=240^\circ$ in Fig.~\ref{fig:deadBand_2c}d, \fm{\edt{the roll-up of the separated shear layer is completely inhibited due to intense BVI in this scenario}. The near-blade vortical field is consequently dominated by fragmented \edt{small flow} structures interacting with the blade rather than by the coherent shear-layer roll-up observed \edt{for the blades with} smaller $H^*$.}

\subsection{Rapid Acceleration Regime}
\label{subsec:res_rapid}

During the rapid acceleration regime, $\omega$ increases rapidly and the flow remains attached to the surface of the blade over a larger portion of the rotation cycle than during the initial acceleration and dead-band regimes. The reduction in the variation of $\alpha_{eff}$ limits flow separation and hinders the formation of a coherent DSV. 


\begin{figure}[h!]
  \centering
  \includegraphics[width=1\linewidth]{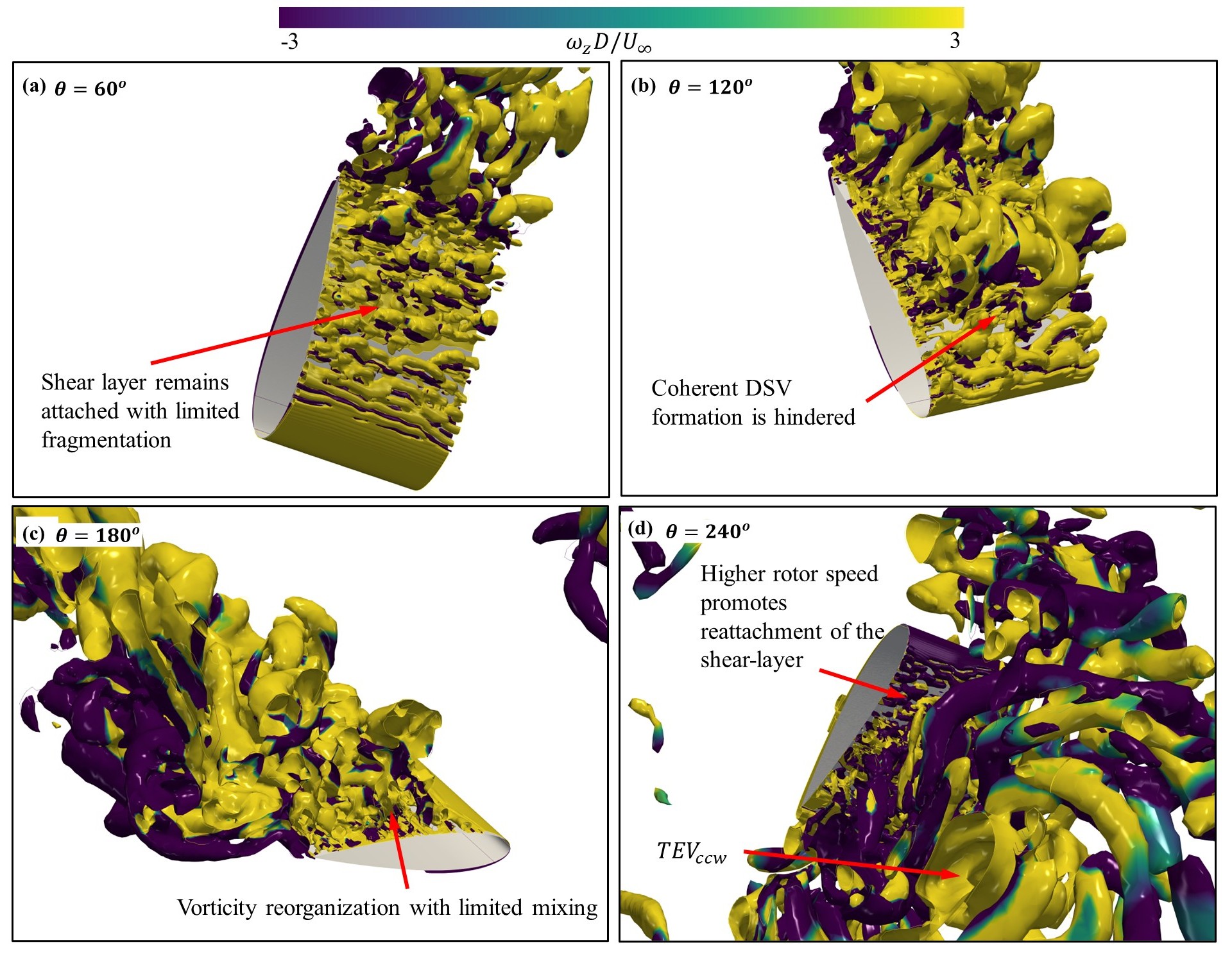}
  \caption{Iso-surfaces of the $Q$-criterion at $Q=10{,}000$ showing the vortex field during the rapid acceleration regime, colored by dimensionless spanwise vorticity $\omega_z D/U_\infty$ for $H^{*}=1$ \fm{figure updated}} 
  \label{fig:rapidAcc_1c}
\end{figure}

For \edt{a blade with} $H^{*}=1$, the shear layer remains predominantly attached to the surface of the blade at $\theta=60^\circ$ in Fig.~\ref{fig:rapidAcc_1c}a, with \edt{some} fragmentation of the vortical structures along the span. This behavior differs considerably from the pronounced flow reversal and roll-up of the shear layer observed at the same azimuthal position during \edt{the stages of} initial acceleration and the dead-band. As the blade advances to $\theta=120^\circ$ in Fig.~\ref{fig:rapidAcc_1c}b, the formation of a coherent DSV is \edt{suppressed}. The separated shear layer does not undergo the extensive roll-up associated with the growth of the large DSV observed during the earlier regimes, and only localized vortical structures remain near the surface of the blade. At $\theta=180^\circ$ in Fig.~\ref{fig:rapidAcc_1c}c, the blade passes through the transitional aerodynamic state between the upwind and downwind half-cycles. The distribution of vorticity reorganizes as the aerodynamic roles of the inner and outer surfaces \edt{are swapped}, while the flow remains considerably more attached than during the dead-band. As the blade progresses to $\theta=240^\circ$ in Fig.~\ref{fig:rapidAcc_1c}d, \fm{the increased \edt{speed of the rotor} enhances the reattachment of the shear layer. The resulting $TEV_{ccw}$, formed as the shear layer reattaches over the inner half of the blade, is observed earlier in the rotation cycle than \edt{that observed} during the initial-acceleration regime, as indicated by its greater downstream convection from the blade.} 

\begin{figure}[h!]
  \centering
  \includegraphics[width=1\linewidth]{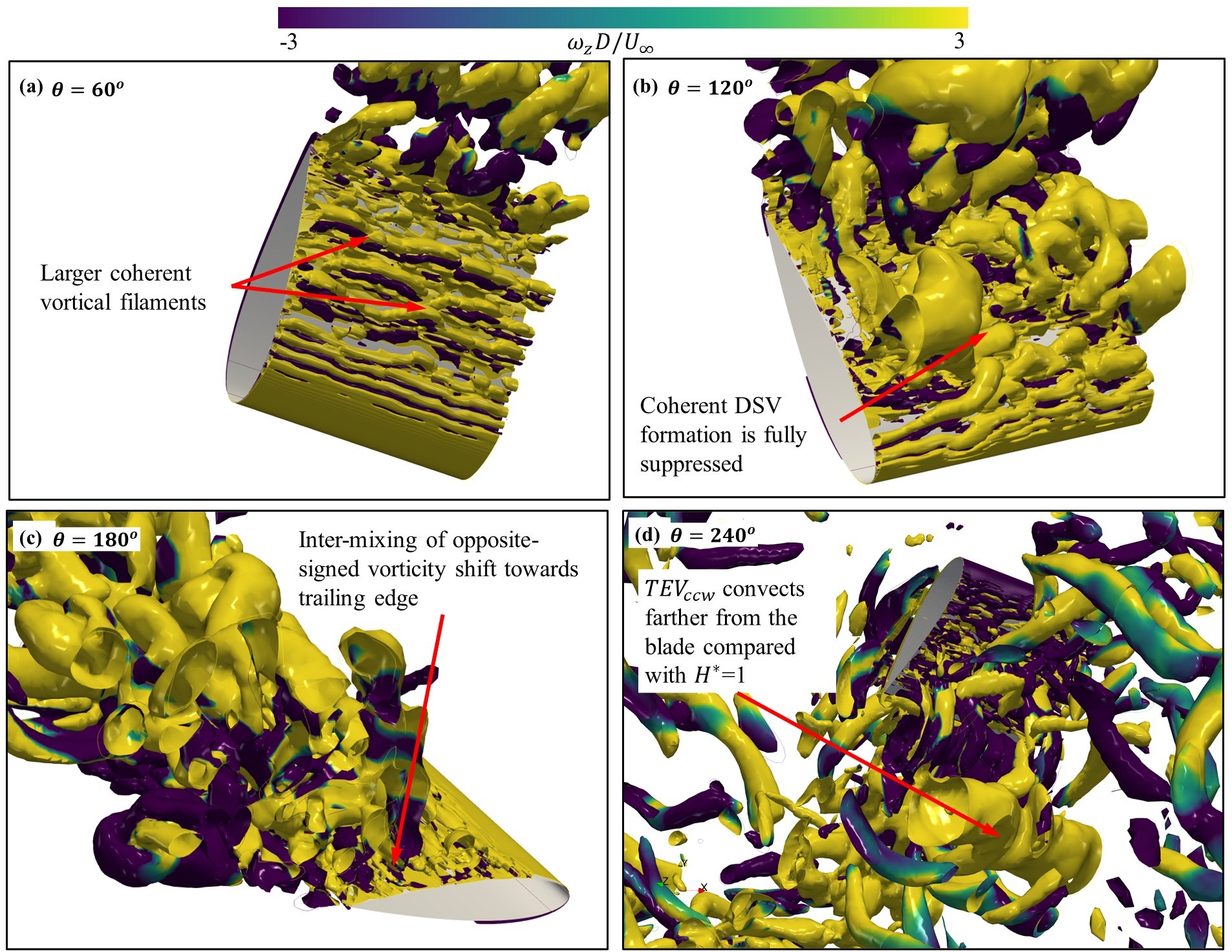}
  \caption{Iso-surfaces of the $Q$-criterion at $Q=10{,}000$ showing the vortex field during the rapid acceleration regime, colored by dimensionless spanwise vorticity $\omega_z D/U_\infty$ for $H^{*}=1.5$ \fm{figure updated}}
  \label{fig:rapidAcc_1.5c}
\end{figure}

\fm{A greater variation of the vortical structures along the span becomes evident as the \edt{span of the blades} is increased to $H^{*}=1.5$, while the overall flow evolution remains similar to that observed for $H^{*}=1$.} At $\theta=60^\circ$ in Fig.~\ref{fig:rapidAcc_1.5c}a, \fm{the shear layer remains predominantly attached, while larger vortical filaments develop along the span.} At $\theta=120^\circ$ in Fig.~\ref{fig:rapidAcc_1.5c}b, the formation of a coherent DSV is fully suppressed and the vortical structures remain confined near the surface of the blade. At $\theta=180^\circ$ in Fig.~\ref{fig:rapidAcc_1.5c}c, \fm{the interaction between \edt{oppositely}-signed vortical structures is confined mainly to the trailing-edge region, consistent with the earlier attachment of the flow as the rotor accelerates.} At $\theta=240^\circ$ in Fig.~\ref{fig:rapidAcc_1.5c}d, \fm{the $TEV_{ccw}$ is observed \edt{to be at a} farther \edt{location} from the blade than \edt{the one} for $H^{*}=1$ and exhibits greater undulation along the span. \edt{Covering a} larger \edt{distance during its shedding and convection} indicates that reattachment of the shear layer over the inner half of the blade, followed by the shedding of $TEV_{ccw}$, occurred earlier in the rotation cycle.}

\begin{figure}[h!]
  \centering
  \includegraphics[width=1\linewidth]{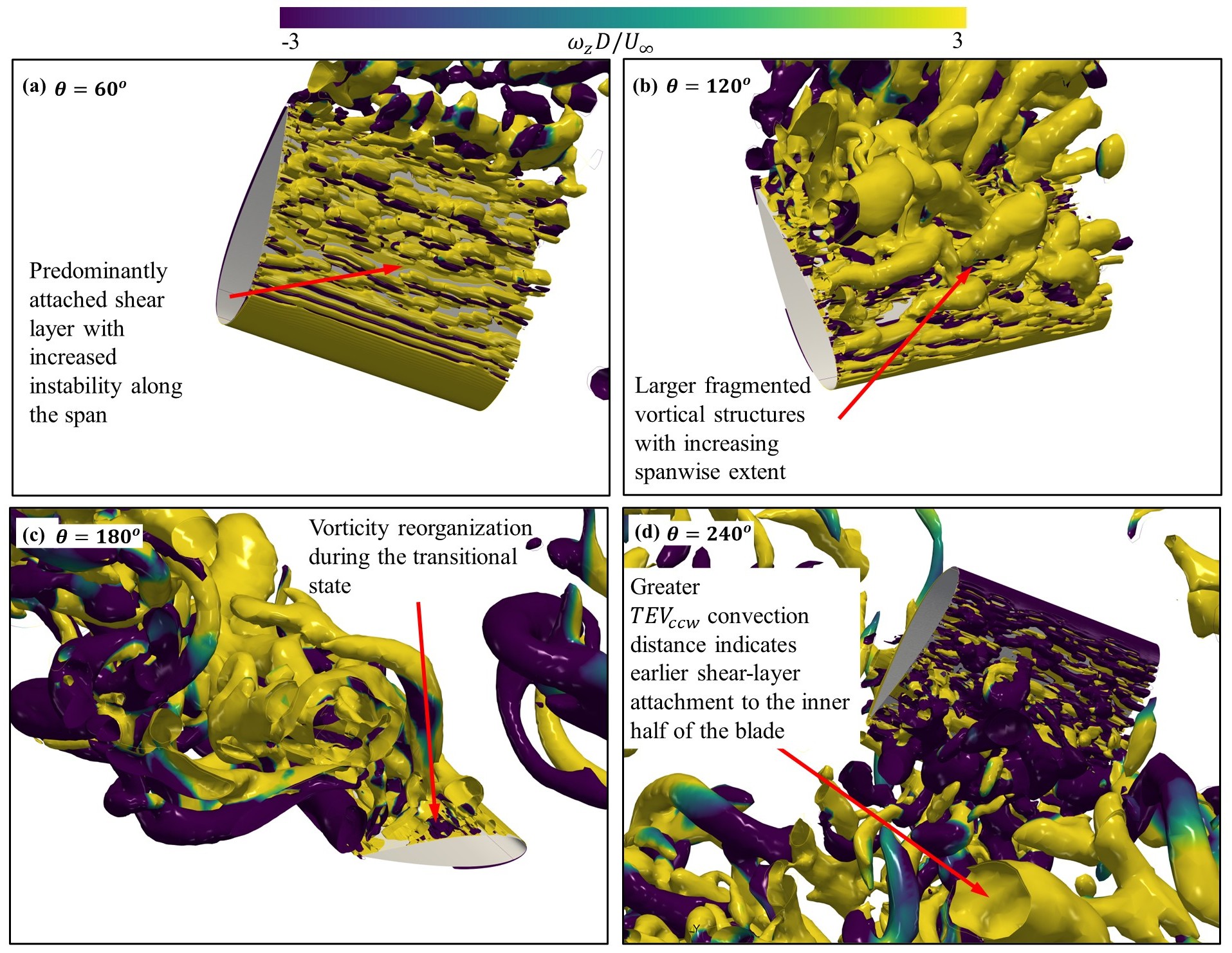}
  \caption{Iso-surfaces of the $Q$-criterion at $Q=10{,}000$ showing the vortex field during the rapid acceleration regime, colored by dimensionless spanwise vorticity $\omega_z D/U_\infty$ for $H^{*}=2$ \fm{figure updated}}
  \label{fig:rapidAcc_2c}
\end{figure}

\fm{\edt{Now, we describe some changes in the overall flow dynamics around the blade with $H^\ast=2$ compared to the previously explained cases in this phase.}} At $\theta=60^\circ$ in Fig.~\ref{fig:rapidAcc_2c}a, the shear layer remains predominantly attached but exhibits greater \fm{instability} along the span, \fm{with the vortical filaments showing increased undulation compared with those observed for the shorter \edt{spans of the blade}.} At $\theta=120^\circ$ in Fig.~\ref{fig:rapidAcc_2c}b, the formation of a coherent DSV remains suppressed, while larger fragmented vortical structures develop near the blade. At $\theta=180^\circ$ in Fig.~\ref{fig:rapidAcc_2c}c, \fm{the \edt{oppositely}-signed vortical structures remain concentrated near the trailing-edge region as the vorticity field reorganizes during the transition between the two half-cycles.} At $\theta=240^\circ$ in Fig.~\ref{fig:rapidAcc_2c}d, \fm{the \edt{vortex} $TEV_{ccw}$ is located even farther from the blade than \edt{that} for $H^{*}=1.5$. \edt{It hints at} a further advancement of the shear-layer reattachment and subsequent vortex shedding within the rotation cycle.}

\subsection{Quasi-Steady Operating Regime}
\label{subsec:res_steady}

In the quasi-steady operating regime, $\omega$ remains nearly constant and the flow around the blade approaches a repeatable periodic state. Compared with the preceding regimes, the flow remains attached to the surface of the blade over a greater portion of the cycle, the formation of a coherent DSV is largely suppressed, and BVI becomes almost negligible. The formation, interaction, and breakdown of the vortical structures are examined for the three values of $H^{*}$ to identify the changes caused by increasing the spanwise extent.

\begin{figure}[h!]
  \centering
  \includegraphics[width=1\linewidth]{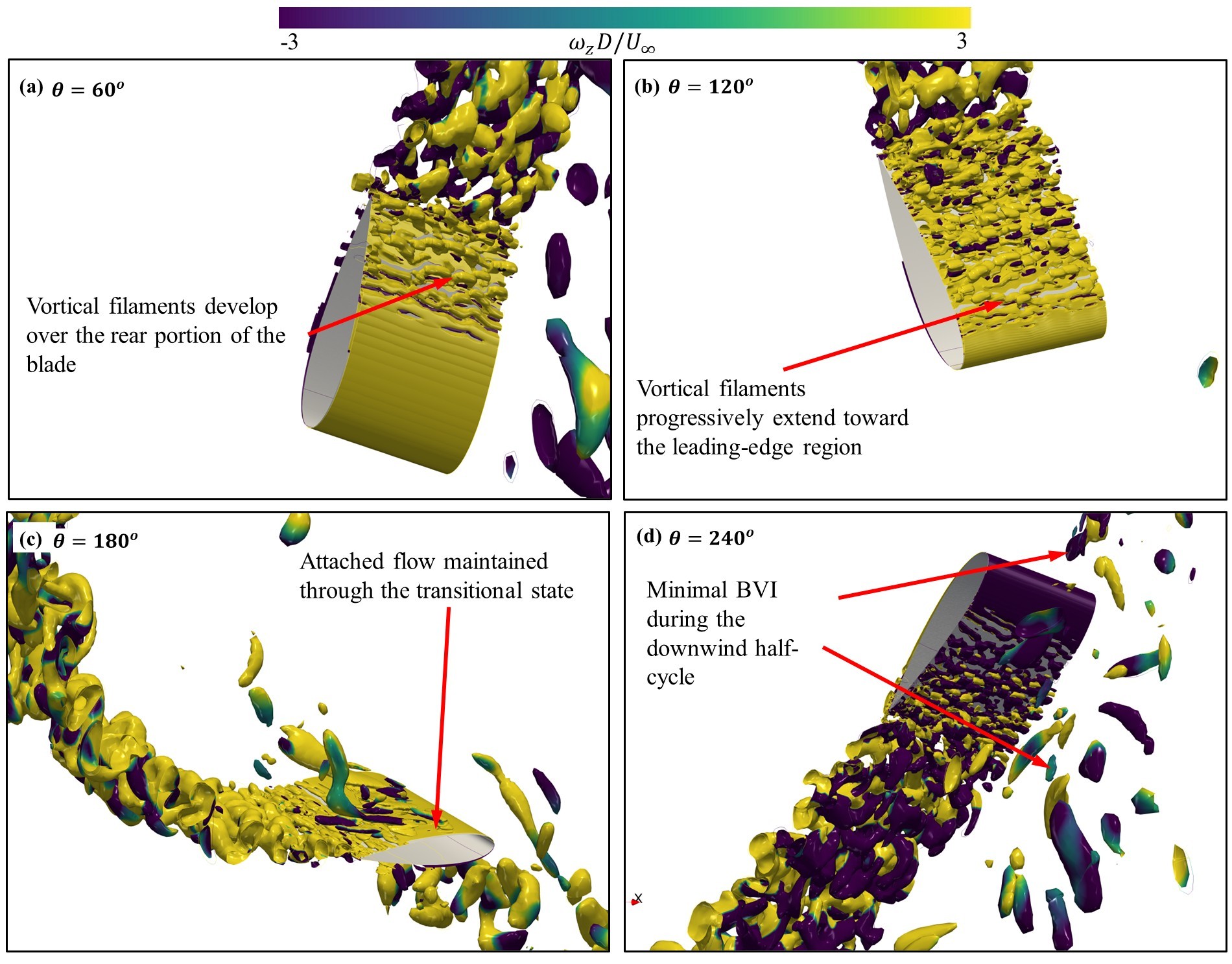}
  \caption{Iso-surfaces of the $Q$-criterion at $Q=10{,}000$ showing the vortex field during the quasi-steady operating regime, colored by dimensionless spanwise vorticity $\omega_z D/U_\infty$ for $H^{*}=1$}
  \label{fig:steadyOp_1c}
\end{figure}

During quasi-steady operation, the flow for $H^{*}=1$ remains predominantly attached at $\theta=60^\circ$ in Fig.~\ref{fig:steadyOp_1c}a, \fm{with vortical filaments developing mainly over the rear portion of the blade and progressively extending toward the leading-edge region as the blade advances to $\theta=120^\circ$ in Fig.~\ref{fig:steadyOp_1c}b.} At $\theta=180^\circ$, \fm{the \edt{interference between} \edt{oppositely}-signed vorticity is no longer prominent, and the flow remains largely attached to the blade. Positive vorticity is confined mainly to the inner surface, while negative vorticity develops over the outer surface. This more organized distribution is associated with the higher speed \edt{of the rotor} attained during the quasi-steady regime.} As the blade progresses to $\theta=240^\circ$ in Fig.~\ref{fig:steadyOp_1c}d, \fm{BVI is minimal because the more attached flow during the preceding upwind half-cycle produces smaller vortices, most of which lose coherence \edt{and gets diffused} before reaching the blade at this azimuthal position.}

\begin{figure}[h!]
  \centering
  \includegraphics[width=1\linewidth]{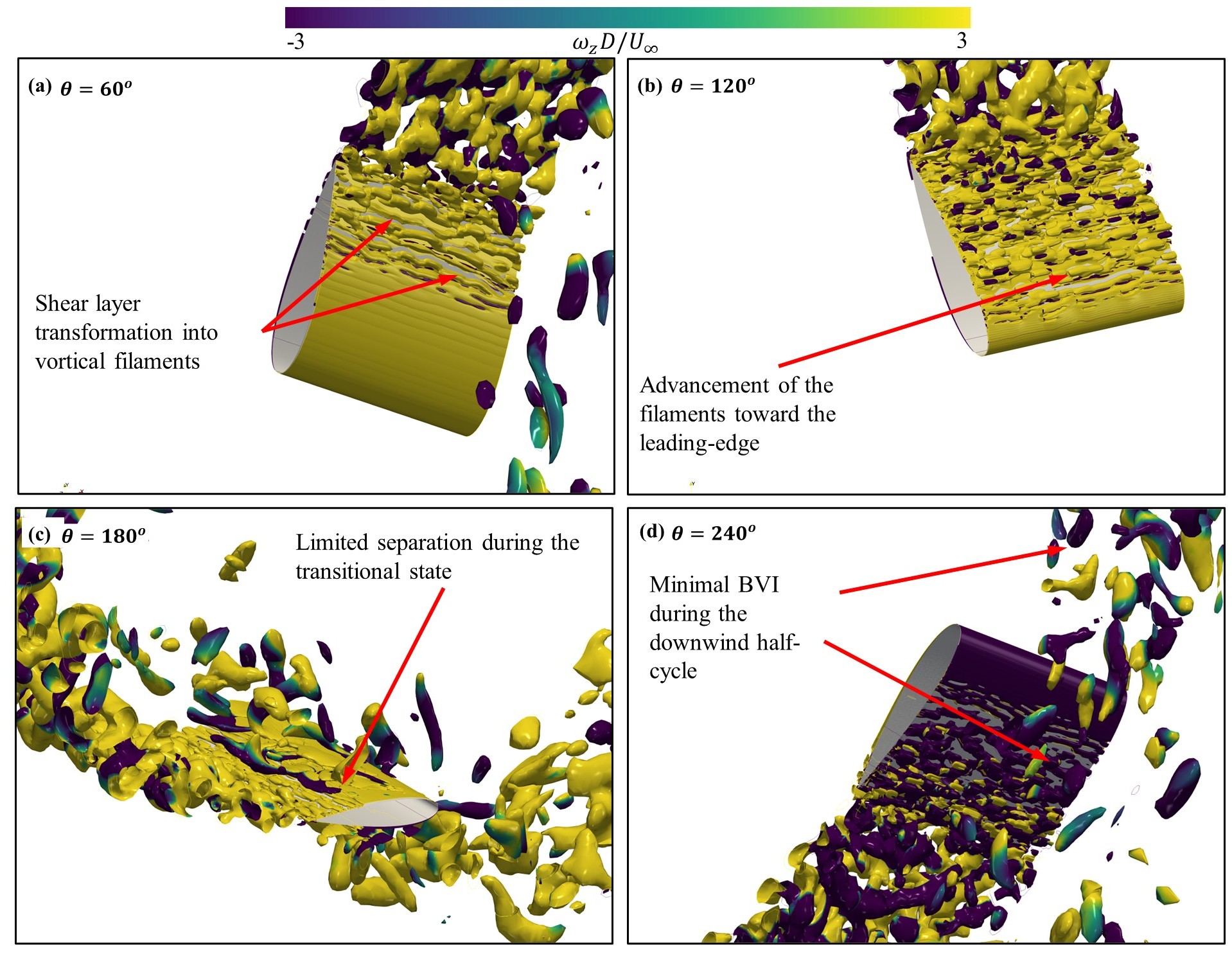}
  \caption{Iso-surfaces of the $Q$-criterion at $Q=10{,}000$ showing the vortex field during the quasi-steady operating regime, colored by dimensionless spanwise vorticity $\omega_z D/U_\infty$ for $H^{*}=1.5$ \fm{figure updated}}
  \label{fig:steadyOp_1.5c}
\end{figure}

\fm{\edt{For the case with $H^{*}=1.5$, we observe some difference in the flow behavior compared to the previous case.}} At $\theta=60^\circ$ in Fig.~\ref{fig:steadyOp_1.5c}a, the flow remains largely attached, with the shear layer transforming into vortical filaments near the trailing edge \fm{that are larger in size and exhibit more pronounced three-dimensional instabilities, and progressively extend toward the leading-edge region as the blade reaches $\theta=120^\circ$ in Fig.~\ref{fig:steadyOp_1.5c}b.} At $\theta=180^\circ$ in Fig.~\ref{fig:steadyOp_1.5c}c, the distribution of vorticity reorganizes during the transitional aerodynamic state, \fm{and minimal BVI is observed as the blade reaches $\theta=240^\circ$ in Fig.~\ref{fig:steadyOp_1.5c}d, indicating that most of the vortices generated during the preceding upwind half-cycle \edt{could not sustain themselves} before interacting with the blade during the downwind half-cycle.}

\begin{figure}[h!]
  \centering
  \includegraphics[width=1\linewidth]{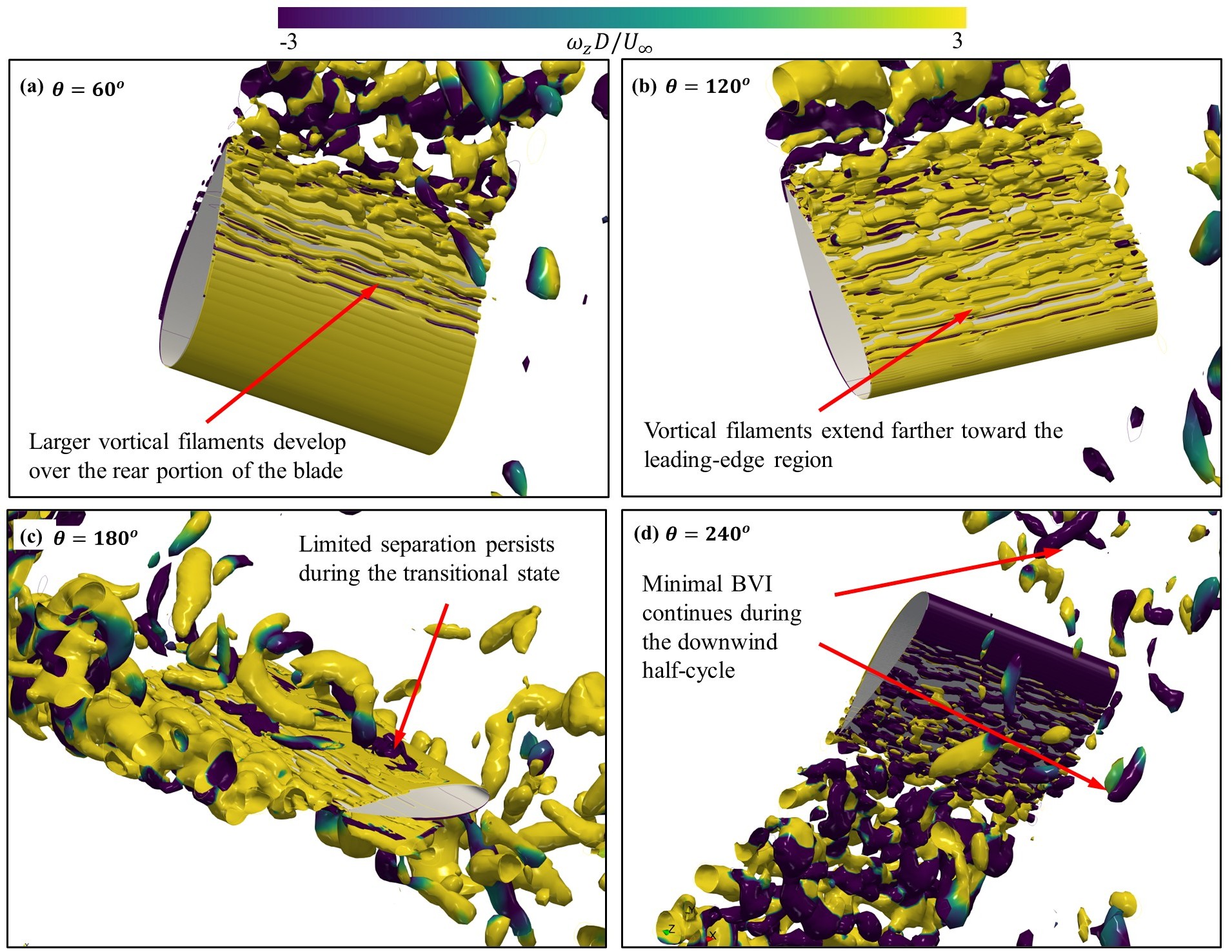}
  \caption{Iso-surfaces of the $Q$-criterion at $Q=10{,}000$ showing the vortex field during the quasi-steady operating regime, colored by dimensionless spanwise vorticity $\omega_z D/U_\infty$ for $H^{*}=2$ \fm{figure updated}}
  \label{fig:steadyOp_2c}
\end{figure}

\fm{\edt{Next, the case with blades having a span of} $H^{*}=2$ retains the largely attached flow observed at the lower computational heights. At $\theta=60^\circ$ in Fig.~\ref{fig:steadyOp_2c}a, larger vortical filaments develop over the rear portion of the blade and extend farther toward the leading-edge region as the blade advances to $\theta=120^\circ$ in Fig.~\ref{fig:steadyOp_2c}b. At $\theta=180^\circ$ in Fig.~\ref{fig:steadyOp_2c}c, limited separation persists during the transitional state, while minimal BVI \edt{is observed,} as the blade reaches $\theta=240^\circ$ in Fig.~\ref{fig:steadyOp_2c}d} with only isolated \edt{small} vortical structures remaining in the vicinity of the blade.

\subsection{Quantitative Assessment of Vortex Development and Flow Attachment}
\label{subsec:quantitative}

\begin{figure}[!t]
\centering
\includegraphics[width=0.98\linewidth]{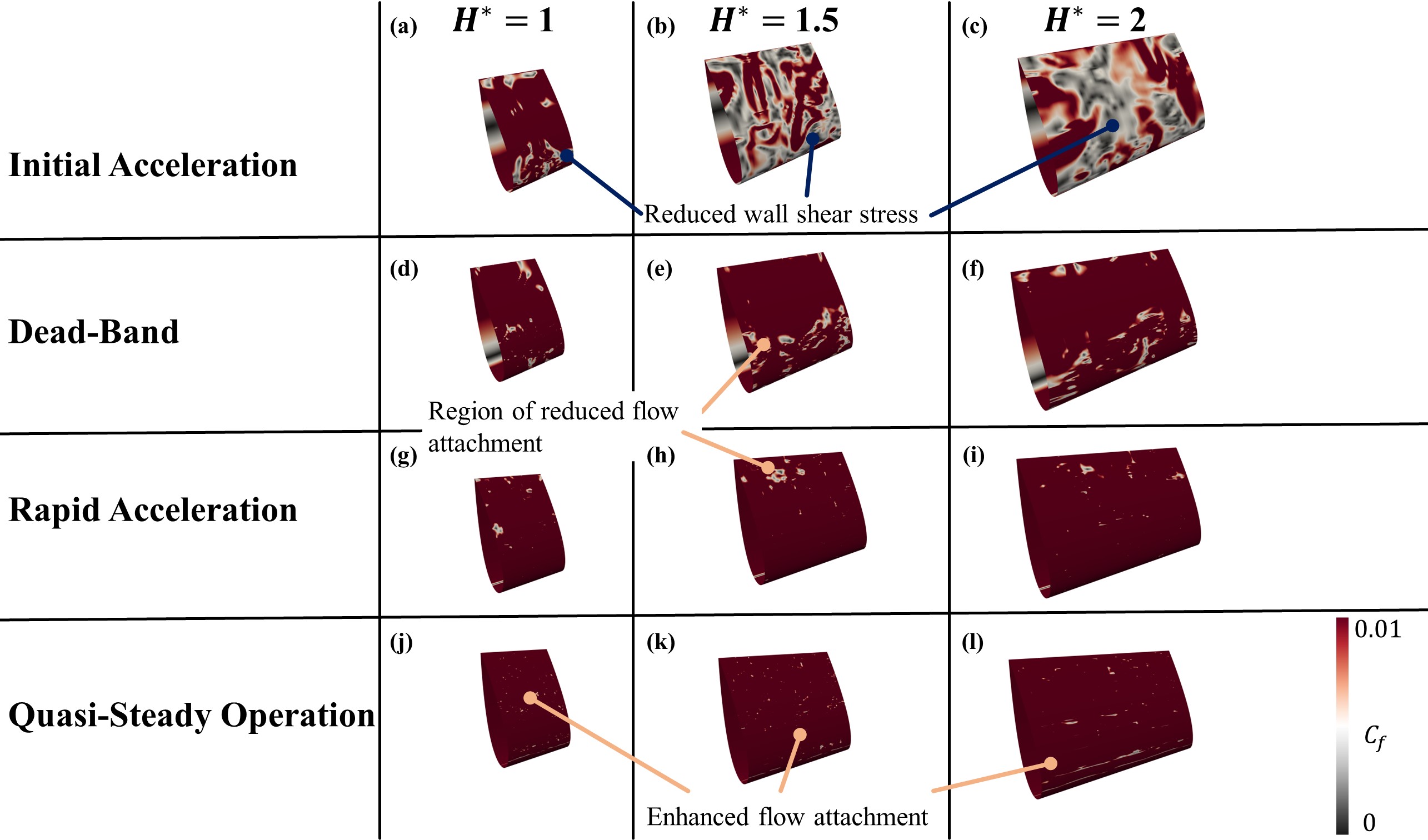}
\caption{Surface distribution of the skin-friction coefficient, $C_f$, at $\theta=120^\circ$ for $H^*=1$, $1.5$, and $2$}
\label{fig:Cf_120}
\end{figure}

\edt{Previously, we provide qualitative explanation of vortical dynamics around a blade during its rotational cycles under different stages of self-starting process of VAWTs. It is important to quantitatively determine the behavior of the flow under these conditions to further examine and elucidate the influence of an increasing span of the blades here. For this purpose,} the \edt{extent} of flow attachment \edt{with the blade} at $\theta=120^\circ$ and $180^\circ$ is assessed using \edt{contours of} the skin-friction coefficient\edt{, and its area-averaged value computed as} $C_{f,a}=(1/A)\int_A C_f\,{\rm d}A$, where $A$ is the blade\edt{'s} surface area. \edt{First, we present} the distributions of $C_f$ at $\theta=120^\circ$ in Fig.~\ref{fig:Cf_120}, and the corresponding values of $C_{f,a}$ are given in Table~\ref{tab:Cf_120}. Regions of low $C_f$, shown by the lighter portions of the contours, indicate lower wall-shear activity and weaker flow attachment. During the initial-acceleration, dead-band, and rapid-acceleration regimes, $C_{f,a}$ decreases with \edt{an} increasing $H^*$, which \edt{explains} weaker flow attachment at larger \edt{spans of the blades} (see Fig.~\ref{fig:Cf_120}a-\edt{\ref{fig:Cf_120}}i). \edt{This} reduction is more pronounced from $H^*=1$ to $H^*=1.5$ ($36.6\%$, $12.6\%$, and $4.3\%$ for the three regimes, respectively) than from $H^*=1.5$ to $H^*=2$ ($1.8\%$, $2.9\%$, and $7.5\%$, respectively), with the largest difference occurring during initial acceleration. \edt{The perviously presented flow visualizations combined with this quantification provide a direct evidence of a higher $H^*$ beneficial for self-starting of VAWTs, because smaller and more deformed flow structures} lose coherence \edt{in these cases} and subsequently does not interact with the blades intensely. However, increasing the \edt{blades' span} of a physical turbine \edt{may} also increase its moment of inertia and could slow its initial acceleration. Therefore, the aerodynamic benefit associated with reduced BVI should be assessed together with the increase in rotational inertia. \edt{Specifically}, it \edt{is} useful to identify the range of $H^*$ over which the largest aerodynamic improvement occurs\edt{,} and beyond which further increases produce only small changes in \edt{the governing vortex dynamics}. 

During quasi-steady operation, the trend changes (\edt{see} Fig.~\ref{fig:Cf_120}j-\edt{\ref{fig:Cf_120}}l), with nearly identical $C_{f,a}$ values for $H^*=1$ and $1.5$ and a higher value for $H^*=2$. \edt{It demonstrates} that the influence of \edt{span of the blades} on \edt{the} flow \edt{dynamics} is dependent on the operating regime. \edt{Nonetheless, it is more important for a VAWT to break through the dead-band phase and reach the quasi-steady rotation. It is important to mention here that variations in $C_{f,a}$ are also dependent on the increased surface area of the blades with an increasing $H^*$.}

\begin{table}[!t]
\centering
\caption{Area-averaged skin-friction coefficient, $C_{f,a}$, at $\theta=120^\circ$ for $H^*=1$, $1.5$, and $2$.}
\label{tab:Cf_120}
\begin{tabular}{lccc}
\hline
Regime & $H^*=1$ & $H^*=1.5$ & $H^*=2$ \\
\hline
Initial acceleration & 0.01712 & 0.01086 & 0.01066 \\
Dead-band            & 0.02668 & 0.02332 & 0.02265 \\
Rapid acceleration   & 0.06063 & 0.05805 & 0.05372 \\
Quasi-steady         & 0.07435 & 0.07416 & 0.08274 \\
\hline
\end{tabular}
\end{table}

\begin{figure}[!t]
\centering
\includegraphics[width=0.98\linewidth]{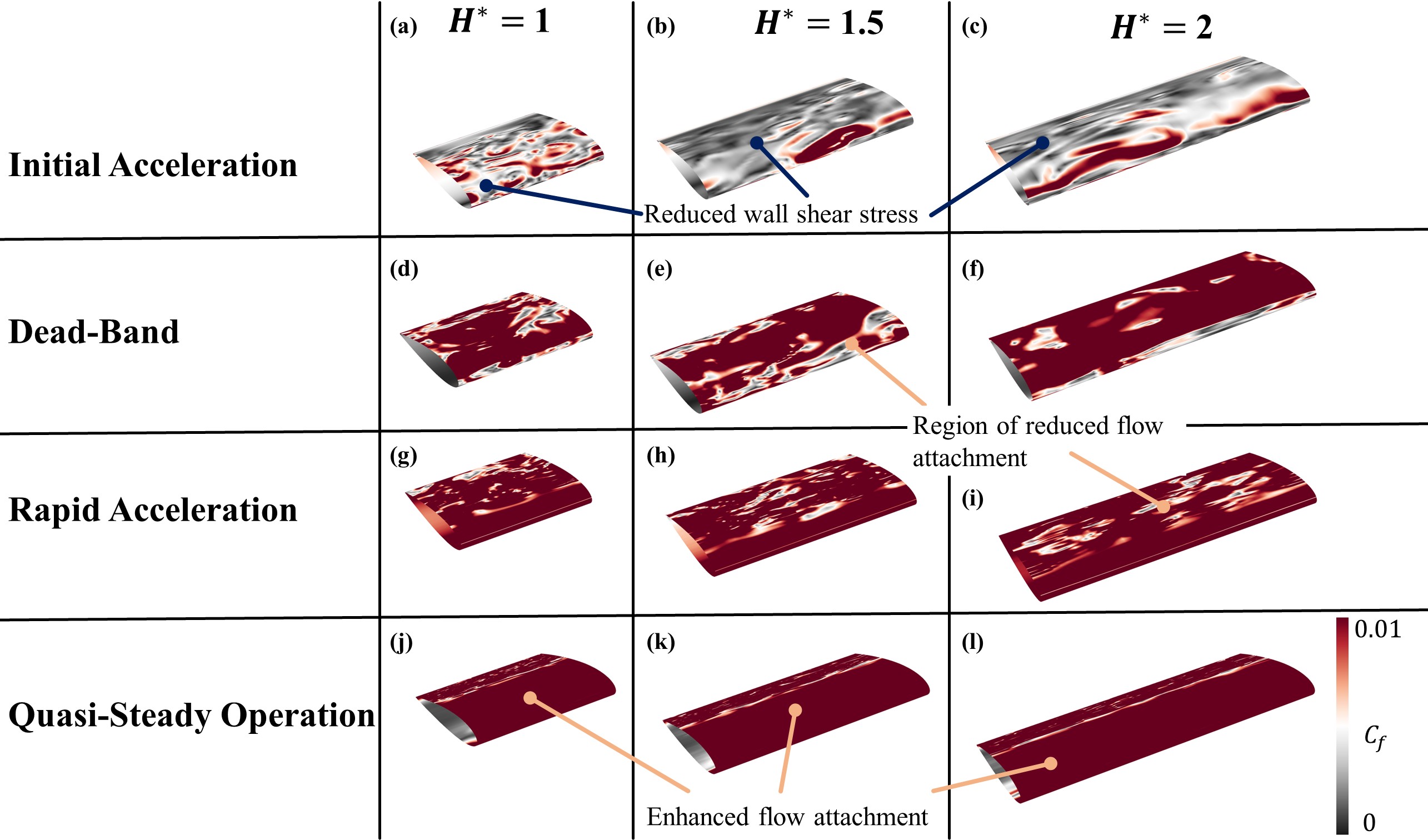}
\caption{Surface distribution of the skin-friction coefficient, $C_f$, at $\theta=180^\circ$ for $H^*=1$, $1.5$, and $2$}
\label{fig:Cf_180}
\end{figure}

The quantitative assessment of $C_{f,a}$ at $\theta=180^\circ$ is based on the surface distributions shown in Fig.~\ref{fig:Cf_180}a-l and the corresponding values listed in Table~\ref{tab:Cf_180}. The values of $C_{f,a}$ are lower than those at $\theta=120^\circ$ for all operating regimes for all values of $H^*$. Unlike the clearer trend observed at $\theta=120^\circ$, the variation here shows no consistent dependence on $H^*$. This behavior is expected because $\theta=180^\circ$ represents the transitional position between the upwind and downwind half-cycles, where the aerodynamic roles of the two blade surfaces interchange and strong \edt{interactions between} opposite\edt{ly}-signed vorticity occurs. As a result, the wall-shear distribution is highly reorganized and does not vary systematically with $H^*$. During initial acceleration, for example, $C_{f,a}$ decreases by approximately $25.2\%$ from $H^*=1$ to $1.5$ and then increases for $H^*=2$, with a similar behavior observed during the dead-band and rapid-acceleration regimes. \edt{Nonetheless, the} increase in $C_{f,a}$ from initial acceleration to quasi-steady operation, indicates that the flow becomes progressively more attached as the rotor accelerates.

\begin{table}[!t]
\centering
\caption{Area-averaged skin-friction coefficient, $C_{f,a}$, at $\theta=180^\circ$ for $H^*=1$, $1.5$, and $2$.}
\label{tab:Cf_180}
\begin{tabular}{lccc}
\hline
Regime & $H^*=1$ & $H^*=1.5$ & $H^*=2$ \\
\hline
Initial acceleration & 0.00453 & 0.00339 & 0.00405 \\
Dead-band            & 0.00917 & 0.00868 & 0.01015 \\
Rapid acceleration   & 0.01880 & 0.01733 & 0.01907 \\
Quasi-steady         & 0.03784 & 0.03886 & 0.03963 \\
\hline
\end{tabular}
\end{table}

\section{Conclusions}
\label{sec:conclusions}

\fm{\edt{In this work, we introduce a new kinematic model through prescribing the rotor's speed, in four stages, as a function of time for examining the aerodynamic behavior and governing three-dimensional vortex dynamics around VAWTs under its self-starting process.} \edt{Using our LES-based computational methodology, we also investigate the influence of the blades' spans on the flow physics of self-starting phenomenon} by using $H^*=1$, $1.5$, and $2$. \fmrev{It is found that increasing $H^*$ leads to a loss of coherence of the DSV, as the separated shear layer remains more attached to the blade surface and consequently does not roll up completely into a coherent vortex.} \edt{This behavior is helpful for the turbine to break through the dead-band phase and reach its quasi-steady rotation.} \edt{We further evaluate the role of flow attachment with the blades of the VAWTs in self-starting process} using area-averaged friction coefficient at two azimuthal positions of a blade. The results show that smaller $H^*$ promotes stronger flow attachment, which is associated with the formation of larger vortical structures during the upwind half-cycle. \fmrev{Larger vortical structures can persist for a longer duration and are more likely to re-encounter the blade during the downwind half-cycle.} These structures can convect downstream and increase BVI \fmrev{more strongly}, adversely affecting the self-starting process. \edt{Besides,} reduction in \edt{the friction coefficient} is much greater from \edt{the case of} $H^*=1$ to \edt{the one for} $H^*=1.5$ than from \edt{those of} $H^*=1.5$ to $H^*=2$. \edt{It highlights that there may be a limit for the span of blades in a VAWT that could contribute substantially in improving its self-starting performance, and beyond which that advantage might not be significant.}} 

\section*{Acknowledgment}
MSU Khalid acknowledges the funding support from Lakehead University through the startup grant and the Natural Sciences and Engineering Research Council of Canada (NSERC) through the Discovery grant program. F. Muhammad is thankful to Lakehead University for the graduate scholarship and for the support through the Ontario Graduate Scholarship. The simulations reported in this work were performed on the supercomputing clusters administered and managed by the Digital Research Alliance of Canada.

\vskip6pt

\bibliography{references}

\end{document}